\documentclass[amsmath,amssym,ams, preprint, superscriptaddress,nofootinbib]{article} 
\usepackage[a4paper,textwidth=17.9cm,textheight=26cm]{geometry}
\usepackage[english]{babel}
\usepackage[explicit]{titlesec}
\renewcommand{\baselinestretch}{1.5} 
\usepackage[T1]{fontenc}
\usepackage{lmodern}
\usepackage{helvet}
\usepackage{lineno}
\usepackage{graphicx}
\usepackage{xr}
\usepackage{authblk}
\usepackage{ulem}
\usepackage{siunitx}
\usepackage{upgreek}
\usepackage{amsmath}

\newcommand{\nocontentsline}[3]{}
\newcommand{\tocless}[2]{\bgroup\let\addcontentsline=\nocontentsline#1{#2}\egroup}

\makeatletter
\newcommand*{\addFileDependency}[1]{
  \typeout{(#1)}
  \@addtofilelist{#1}
  \IfFileExists{#1}{}{\typeout{No file #1.}}
}
\makeatother

\addto\captionsenglish{}
\addto\captionsenglish{}

\begin{document}

\title{Wireless millimeterwave electro-optics on thin film lithium niobate}
\author[1, 2]{Aleksei Gaier}
\author[1]{Karen Mamian}
\author[3, 4]{Shima Rajabali}
\author[1, 2]{Yazan Lampert}
\author[1, 2]{Jiawen Liu}
\author[3]{Leticia Magalhaes}
\author[3, 5]{Amirhassan Shams-Ansari}
\author[3]{Marko Lončar}
\author[1, 2]{Ileana-Cristina Benea-Chelmus}

\affil[1]{Hybrid Photonics Laboratory, École Polytechnique Fédérale de Lausanne (EPFL),  CH-1015, Switzerland}
\affil[2]{Center for Quantum Science and Engineering (QSE), CH-1015, Switzerland}
\affil[3]{Harvard John A. Paulson School of Engineering and Applied Sciences, Harvard University, Cambridge, MA, USA}
\affil[4]{Department of Quantum and Computer Engineering, Delft University of Technology, Netherlands}
\affil[5]{DRS Daylight Solutions, 16465 Via Esprillo, CA, USA}

\date{\today}
\maketitle


\begin{abstract}
The rapid growth of global data traffic is accelerating the need for ultra-broadband communication technologies, particularly in cloud infrastructure and emerging 6G wireless systems. Optical computing and quantum information processing also demand fast, scalable ways to interface optical and electronic signals. Integrated electro-optic modulators provide a compact and efficient solution, but extending their operation into the millimeterwave (mmWave) range with wide bandwidth and compatibility with wireless signals remains a significant challenge. Bulky electrical packaging and high mmWave losses remain primary barriers to scalability and high performance.

Here, we demonstrate a wireless and wideband electro-optic modulation architecture that directly interfaces mmWaves with optical signals, eliminating the need for impedance-matched mmWave probes and cables. By integrating an on-chip antenna with a co-designed transmission line on thin-film lithium niobate platform, we achieve wideband modulation across the WR9.0 (82–125 GHz) and WR2.8 (240–380 GHz) bands. The wideband nature of our modulator enables the device to function as a high-speed detector of mmWave carriers modulated up to 6~GHz and achieves a flat and wide response, a key requirement for 6G and high-speed mmWave sensing. By configuring the antenna-coupled transmission line to operate as a mmWave cavity, our wireless platform enables triply resonant electro-optic transduction, supporting frequency comb generation with mode spacing of 123.2 GHz and 307.9 GHz. Extracted single-photon electro-optic coupling rates of  $g_{0} =2\pi\times 4.98$~kHz and $2\pi\times9.93$~kHz at 123.2 GHz and 307.9 GHz, respectively, highlight favorable scaling with increasing mmWave frequency. These results introduce a new class of wireless electro-optic devices for high-speed modulation, detection, and frequency comb generation, with impactful applications in communications, sensing, and quantum technologies.
\end{abstract}

\maketitle

\tocless\section{Introduction}\label{sec_intro}
The rapid expansion of cloud services, AI workloads, and global data exchange has led to an unprecedented surge in data traffic, placing stringent demands on the underlying communication infrastructure~\cite{tauber2023role}. Exploiting under-utilized frequency bands such as the mmWave (30-300 GHz) and the terahertz (300 GHz - 10 THz) offers a promising path to meet the demand for this ever-increasing growth~\cite{HUANG2023106}. These bands offer broader bandwidths than  traditional microwave frequencies and  exhibit improved atmospheric robustness compared to free-space telecom-wavelength optical links. 
However, all-electronic transmitters, receivers and mixers needed to establish a mmWave/terahertz communication link, typically based on high-frequency electronic devices such as high-electron-mobility transistors~\cite{haziq2022challenges} or Schottky barrier diodes~\cite{kou2022review}, face fundamental challenges. In particular, the signal integrity and the bandwidth of these systems are limited due to technical issues. 
For example, a typical way to detect high-frequency signals is to implement heterodyne detection using a local oscillator. Since this local oscillator must be at mmWave frequencies, it's often generated using frequency extenders~\cite{kou2022review}. These components generate several harmonics of the fundamental frequency, but they suffer from insufficient suppression of undesired harmonics, compromising the purity of the generated signal~\cite{harter2020generalized} and consequently also of the detected mmWave.
Similarly, the bandwidth of the intermediate frequencies (IF) of commercial all-electronic mixers is about a few tens of gigahertz, limiting the data modulation (demodulation) speed at the sender (receiver) end~\cite{jia20202}. Moreover, high power consumption required for signal generation and detection, along with susceptibility to electrostatic discharge impose further limitations on all-electronic systems~\cite{meng2022mscl}.

Photonic technologies offer a promising solution to these challenges~\cite{wang2024ultrabroadband, horst2025ultra, zhu2025integrated}. A key component that overcomes the aforementioned problems at the receiver end is an integrated electro-optic modulator, which converts microwave signals to the optical domain within a compact and power-efficient chip. The interaction between microwave signals and optical carriers grants access to wide bandwidths, reduced propagation losses over long distances by using fiber technologies, and signal multiplexing. This domain, known as microwave photonics, has already proven impactful in high-speed communication systems~\cite{zhang2021integrated} and sophisticated sensing platforms~\cite{zhu2025integrated}.
Of particular interest for next generation communication systems are wireless modulators which are designed to directly capture free-space signals and encode them onto optical carriers~\cite{HUANG2023106}. By providing a direct interface between the wireless signals used in communications systems such as 5G, with optical data transmission infrastructure, these modulators enable high-capacity links between wireless access points and long-haul fiber networks.

To further meet the demands of next-generation technologies such as the emerging 6G networks, there is a growing interest in extending the operation of electro-optic modulators from traditional microwave frequencies into the millimeter-wave (mmWave) and terahertz regimes~\cite{HUANG2023106}.
Among various mechanisms explored for phase and intensity modulation, electro-optics stand out due to its intrinsically fast electronic response, enabling modulation bandwidths that far exceed the limitations of electro-absorption modulators.
High-speed modulation in mmWave and terahertz spectral bands would not only benefit optical communication but also enable critical advances in quantum optics~\cite{santamaria2018sensitivity, zhu2022spectral, yamaguchi2023advanced, couture2025terahertz}, quantum transduction between optical and microwave domains~\cite{multani2025integrated, holzgrafe2020cavity, mckenna2020cryogenic}, mmWave radar and ranging systems~\cite{yi2023photonic, zhu2025integrated}, and quantum computing architectures~\cite{anferov2024millimeter, kumar_quantum-enabled_2023, Suleymanzade}. 

Achieving efficient light modulation at mmWave and terahertz frequencies, while maintaining compact, integrated, and low-loss device architectures, remains a key technological challenge. This has driven extensive research across various material platforms~\cite{Rajabali2023Present}. Material platforms under investigation include lithium niobate~\cite{hu2025integrated, wang2018nanophotonic, mercante2018thin, xu2020high, zhang2021integrated, xie2025broadband}, hybrid silicon nitride–lithium niobate~\cite{zhang2021high, badri_compact_2025}, hybrid silicon-organic modulators~\cite{salamin2015direct,  salamin2019compact, witmer2020silicon, horst2025ultra}, aluminum nitride~\cite{xiong2012low, zhu2016aluminum, liu2023aluminum}, barium titanate~\cite{rosa2017barium, han2023integrated} and lithium tantalate~\cite{wang2024ultrabroadband}.

When selecting a platform for mmWave and terahertz applications, two properties must be considered: the terahertz loss, and the broadband electro-optic nonlinearity of the material. Aluminum nitride is favorable in terms of terahertz loss owing to its high frequency optical phonons~\cite{majkic2015optical}.  However, its relatively low electro-optic coefficients ($r_{33} = -0.59~$pm/V) limits the modulation efficiency~\cite{graupner1992electro}. In contrast, barium titanate has exceptionally high Pockels coefficients (${r_{42}}= 500~$pm/V), but suffers from large terahertz loss~\cite{WAN20082137}, and a rapid rolls-off in nonlinearity beyond 80~GHz~\cite{chelladurai2025barium}. Organic materials exhibit low terahertz loss and have demonstrated record-breaking modulation bandwidths~\cite{horst2025ultra}. However, their high optical absorption limits their power handling, and they currently do not constitute a stand-alone photonic platform, requiring co-integration with other materials~\cite{wang2023perspectives, benea2021electro}. 
Thin film lithium niobate~(TFLN) offers both low optical loss, moderate nonlinearities and high-power stability. Recently, TFLN enabled both broadband terahertz generation~\cite{herter2023terahertz, lampert2024photonics}, terahertz detection and beam profiling at 500~GHz~\cite{tomasino2024large}, and on-chip monolithic electro-optic modulation and terahertz generation~\cite{zhang2024monolithic} as early demonstrations of hybrid terahertz-photonic devices. 

A critical but relatively underexplored challenge in developing mmWave electro-optic modulators is efficiently delivering the mmWave signal to the chip. Currently, most demonstrations rely on probe-based coupling methods, which become increasingly fragile, expensive, and difficult to scale at frequencies above a few tens of gigahertz. These methods suffer from significant attenuation of high-frequency signals (scaling with $\propto \sqrt{f_{mmW}}$~\cite{gallagher1987subpicosecond, keil1992high}),  due to decreasing skin depth. Both the electrical circuit delivering the mmWave signals to the photonic chip and circuit on the chip contribute to this loss. Another complication arises from increased radiative loss of on-chip transmission line which scales as $\propto w^2 f_{mmW}^3$~\cite{keil1992high, martin2015artificial}, with $w$ being the distance between ground and signal electrodes. This radiative loss leads to elevated cross-talk at higher frequencies. Additionally, impedance matching between the coaxial cables, the probes, and the on-chip transmission line (typically 50 $\Omega$) presents a significant challenge in mmWave and terahertz frequency range. We note that the current mmWave delivery technology relies on rectangular waveguides to mitigate radiative and metallic loss. However, these waveguides are bulky and lack the necessary flexibility for integration.

These challenges highlight the need for developing practical and low-loss techniques to enhance the efficiency of electro-optic modulation across the entire signal path. Transitioning from wired to wireless approaches offers a promising solution by transmitting millimeterwave signals through free space, thereby avoiding ohmic losses, and directly coupling them into the chip via large aperture antennas, eliminating the need for complex interfaces. Previous demonstrations of this approach~\cite{salamin2015direct, zhu2025integrated, murata2021antenna, murata2020millimeter} have been restricted to a relatively narrow operational bandwidth.

\begin{figure}[h!]
    \centering
     \includegraphics[width=18cm]{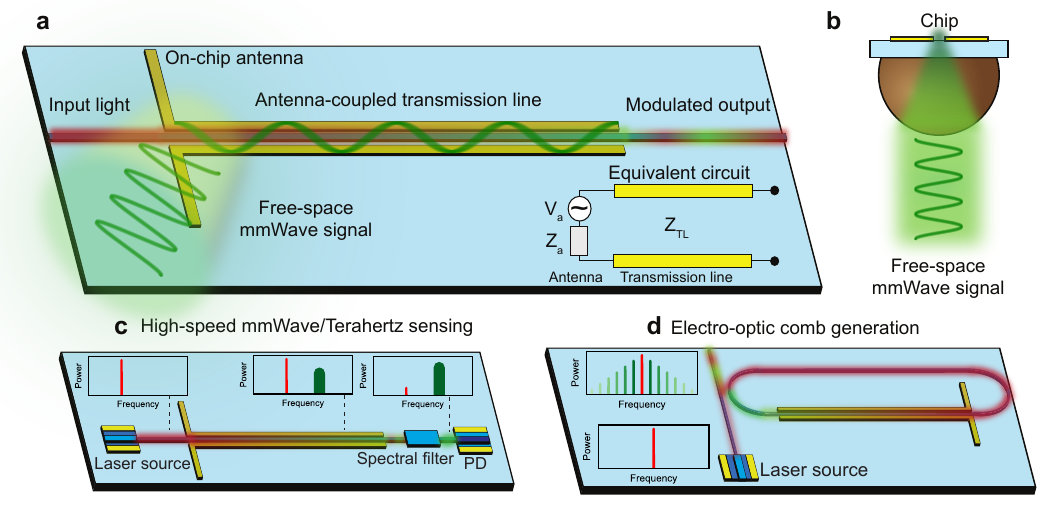}
    \renewcommand{\baselinestretch}{1} 
    \caption{\textbf{Wireless mmWave electro-optics (EO) on thin film lithium niobate (TFLN) with applications in wideband modulators, high-speed sensing and mmWave frequency comb formation.} (a) Conceptual sketch of a wireless electro-optic (EO) modulator, where a free-space mmWave signal is captured by an on-chip antenna and coupled to a transmission line. This transmission line is fabricated along a TFLN-based optical waveguide. The captured mmWave modulates the phase of the light through the electro-optic effect. (b) Side-view illustration of the wireless signal delivery scheme. An incoming mmWave or terahertz beam is focused onto the on-chip antenna using a hyper-hemispherical silicon lens to maximize coupling efficiency. 
    (c) Schematic of a high-speed free-space mmWave detector. A laser pumps the optical waveguide, and incident mmWave or terahertz radiation induces phase modulation in the interaction region along the optical path, generating an optical sideband in the frequency domain. The filter enables the detection of sidebands on a photodiode (PD). (d) Schematic of a wireless and mmWave spacing EO comb frequency comb generator. An incoming wireless signal phase-modulates the input optical signal, resulting in the generation of multiple sidebands as the light undergoes several roundtrips within the cavity.}
    \renewcommand{\baselinestretch}{1.5} 
    \label{fig1}   
\end{figure}

In this work, we propose a wireless and wideband mmWave and terahertz electro-optic modulator on TFLN platform that operates across the WR9.0 (82-125~GHz) and WR2.8 (240-380 GHz) bands (Fig.~\ref{fig1}~a). Our device implements an on-chip antenna coupling free-space terahertz radiation to an on-chip transmission line eliminating the need for complex and fragile terahertz signal delivery components. To ensure the broadband operation of this device, the phase index of the mmWave traveling across the transmission line is engineered to match the group index of the optical signal enabling velocity matching over a wide frequency range. Furthermore, we integrate a hyper-hemispherical silicon lens on the backside of our TFLN chip (aligned with the on-chip antenna) to improve the coupling efficiency of the incoming mmWave/terahertz signal (Fig.~\ref{fig1}~b).
Upon being illuminated by the external mmWave source, the antenna becomes a mmWave signal generator with a voltage amplitude $V_a$ and characteristic impedance $Z_a$ that can be tailored through design (inset of Fig.~\ref{fig1}~a). This flexibility significantly simplifies the co-design of the transmission line to minimize optical and mmWave losses while maximizing the modulation efficiency. 
Conventional characterization techniques are incompatible with miniaturized mmWave devices, necessitating novel approaches to extract key transmission line parameters. To address this, we develop a photonics-based characterization method that shows excellent agreement between theory and experiment. This technique provides a critical step toward verifying the material properties of thin films relative to their bulk counterparts at mmWave frequencies.

Leveraging our flat frequency response we realize a miniaturized high-speed terahertz detector (Fig.~\ref{fig1}~c). In contrast to the state-of-the-art terahertz detectors that rely on pyroelectric effect limiting their speed to a few kHz~\cite{wubs2024performance}, our device operates 6-orders of magnitude faster, reaching several gigahertz range. This dramatic improvement stems from its operation based on the ultra-fast electro-optic effect with its detection speed currently limited by the modulation bandwidth of the incoming signal. Finally, we leverage the low-loss nature of TFLN platform and place our phase modulator in a racetrack cavity (Fig.~\ref{fig1}~d). This enhances the effective path length over which the optical signal interacts with the mmWave, and results in a cascaded sideband generation. Furthermore, designing the antenna-coupled transmission line as a mmWave resonator, we exploit the triply resonant electro-optic transduction to realize a proof-of-principle electro-optic frequency comb source with adjustable mmWave line spacings of 123~GHz and 308~GHz.

 Altogether, our work lays the theoretical and experimental foundation for designing wireless terahertz modulators with smaller footprints, enabling scalability, broad bandwidths, and outlines strategies to achieve higher efficiencies through resonant nonlinear effects. 

\tocless\section{Results}\label{sec_results}
\subsection*{Antenna-transmission line co-design rules}\label{subsec_antennamodelling}
 Similar to conventional modulators, for our wireless modulator we need to ensure that the mmWave signal on-chip is maximized, while optimizing its interaction with the optical carrier. The antenna design dictates the amount of power which is collected from free space and delivered to the transmission line, while the transmission line design impacts the modulation efficiency between the optical and the mmWave signal. Our theoretical model provides the following design guidelines:

i) \textbf{The transmission line:} 
The geometry of the transmission line must be engineered so it maintains a coherence length ($L_{coh}$) larger than the device length $L_{TL}$, minimizes mmWave propagation loss $\alpha_{\Omega}$, maximizes spatial overlap $\Gamma_{eo}$ between the optical and mmWave modes, and achieves strong mmWave mode confinement by reducing the mode cross-section $S_\Omega$. However, unlike conventional modulators, our wireless modulator does not require the impedance matching between the characteristic impedance of the transmission line $Z_{TL}$ and 50 ohms of cables and delivery circuitry.

ii) \textbf{The antenna:} The antenna must
be optimized to collect the maximum amount of mmWave signal (with free-space field amplitude $E_{inc}$). Upon incidence, the antenna generates a voltage wave $V_{a}$ across its gap. The goal is to maximize the frequency-dependent inverse antenna factor ($IAF=\frac{V_{a}}{E_{inc}}$). 
Additionally, ensuring impedance matching between the antenna and the transmission line ($Z_a = Z_{TL}$) leads to a maximum delivery of the collected wave to the transmission line. An impedance mismatch ($Z_a \neq Z_{TL}$) results in a partial reflection at the antenna-transmission line interface, leading to the formation of a Fabry-Perot type mmWave cavity.

\subsection*{Wideband and wireless mmWave modulator}
To meet the design rules, we optimize two critical transmission line parameters, namely, the electrode gap ($w = 3.3~\mu$m), and the optical waveguide width ($w_{wg}=1.5~\mu$m) to achieve phase matching, minimize the metal-related absorption, and maximize the coherence length. We achieve an optical loss of 1 dB/mm, mode overlap factor of $\Gamma_{eo} = 0.54$, and coherence length of $L_{coh}=5~$mm with the aforementioned values for the waveguide width and electrode gap. We note that the measured optical loss is higher than state-of-the-art values~\cite{wang_integrated_2018}, however we chose this narrower gap to achieve a small mode area and hence improve the modulation, highlighting the importance of balancing these parameters rather than minimizing the loss alone.

 On the antenna's side, we opt for a dipolar design as opposed to bow-tie antennas since its resonant behavior will boost the $IAF$.  
 However, this comes at the cost of a frequency-dependent impedance $Z_a$~\cite{balanis2016antenna} which complicates the impedance matching with the transmission line, leading to reflection of the incident mmWave signal at the transmission line interface (see details in the Supplementary Information section \ref{ReflmodelSI}). Our simulations confirm that an antenna design with a length of 400 $\mu$m gives high values of $IAF$ at frequency range 50-500 GHz and a frequency-dependent reflection coefficient $r_a$ varying in absolute value between 0.2-0.9. Further details in Supplementary Information section \ref{suppl:coupled_THz_fields} and \ref{CouplEffStrategiesSI}. 

\begin{figure}[h!]
    \centering
    \includegraphics[width=18cm]{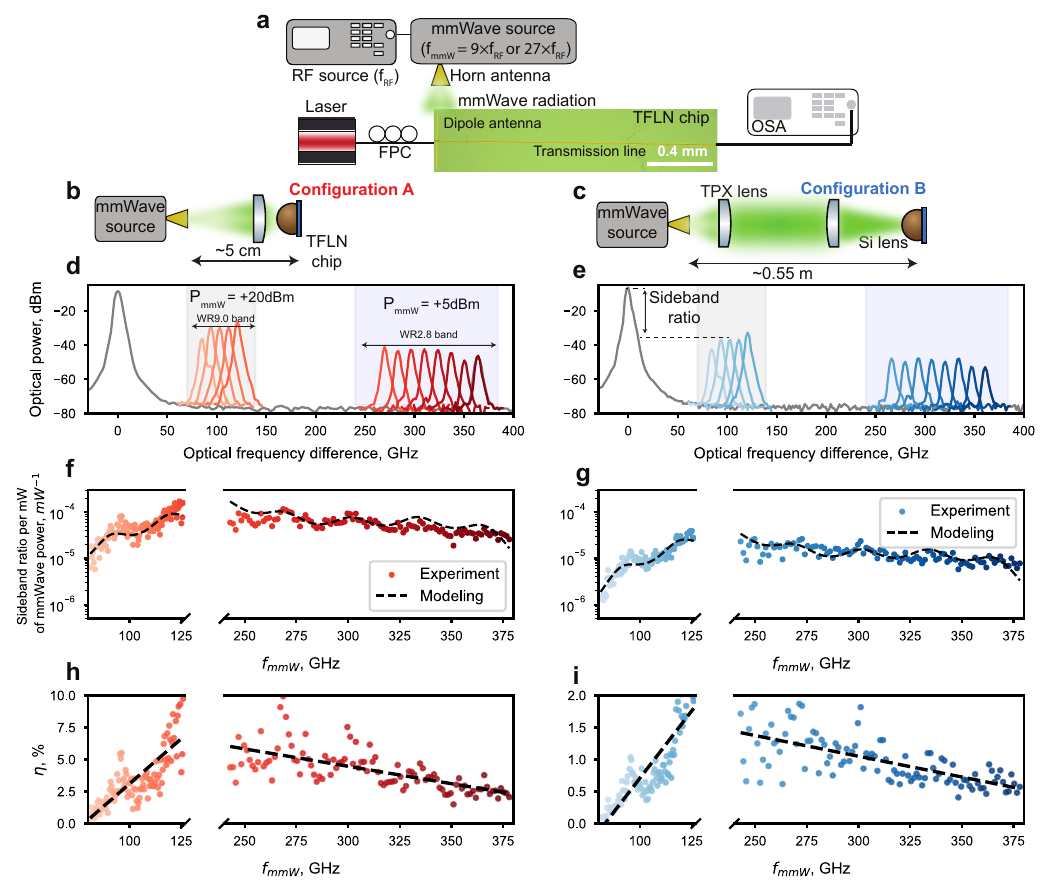}
    \renewcommand{\baselinestretch}{1} 
    \caption{\textbf{TFLN-based wideband mmWave wireless phase modulator.} (a) Schematic of the setup: an RF source generating continuous-wave microwave signal at frequencies between 9–14 GHz drives a frequency chain multiplier (mmWave source) with a multiplication factor of 9 (WR9.0 band) or 27 (WR2.8 band). The resulting mmWave signal is emitted into free space via a horn antenna and couples to mmWave transmission line on-chip. The mmWave phase-modulates the continuous-wave light in the TFLN waveguide, leading to the formation of optical sidebands. The output spectra are recorded on an optical spectrum analyzer (OSA). Microscope image of the fabricated chip is shown as a part of the measurement setup. (b) Configuration A (short range illumination): The mmWave beam is focused onto the chip using a single TPX (polymethylpentene) lens, with a total distance of 5 cm from the horn antenna. Example applications includes mmWave delivery in a cryostat. (c) Configuration B (medium range illumination) corresponds to a setup with. The total distance of 0.55 m between the chip and horn antenna. In both cases we attach a hyper-hemispherical silicon (Si) lens to the back-side of the TFLN chip to improve the coupling efficiency. (d) Measured optical spectra (offset from the optical carrier) for various mmWave frequencies for Configuration A. (e) Measured optical spectra (offset from the optical carrier) for various mmWave frequencies for Configuration B. (f) Sideband ratio per mW of incident mmWave power as a function of frequency in Configuration A. (g) Sideband ratio per mW of incident mmWave power as a function of frequency in Configuration B. (h) Coupling efficiency ($\eta$) as a function of frequency for Configuration A, calculated from eq.\ref{mod-eff-eq}, with a linear fit shown as a dashed line. (i) Coupling efficiency ($\eta$) as a function of frequency for Configuration B, with linear fit shown as a dashed line. FPC – fiber polarization controller. OSA - optical spectrum analyzer.}
    \renewcommand{\baselinestretch}{1.5} 
    \label{fig2}   
\end{figure}

The devices are fabricated in an x-cut thin film lithium niobate substrate similar to our previous work~\cite{lampert2024photonics,tomasino2024large}. The transmission lines are  $L_{TL} = 2~$mm 
long which is limited by mmWave loss. We illuminate our devices with free-space mmWave radiation from a frequency extender system equipped with horn antennas emitting a Gaussian mmWave beam (Fig.~\ref{fig2}~a). We investigate two scenarios: configuration A - short range ($< 10$~cm) to simulate signal delivery in a constrained space such as inside a cryostat for quantum experiments and configuration B - medium range ($10-100$~cm) for applications in open spaces, such as 6G communications (shown in Fig.~\ref{fig2}~b). Details on the fabrication and the experimental setup in the methods and Supplementary Information section \ref{SIstructure}. 

We butt-couple 1550 nm continuous wave light to the chip (with an insertion loss of 9 dB/facet). The polarization of the light is aligned with the z-axis of lithium niobate (TE). To ensure maximum modulation efficiency, and benefit from the largest component of the nonlinear susceptibility tensor $\chi^{(2)}_{333}$, the antenna collects the free-space radiation into the plane of TFLN (z-axis), and couples it to the transmission line, generating an electric field parallel to the polarization of the optical field.
In the frequency domain, the phase modulation (three-wave mixing) of the optical carrier by the mmWave is represented by the generation of two optical sidebands. We quantify the modulator's efficiency using the sideband ratio ($SBR$) at a given mmWave power incident on the chip, $P^{free-space}_{mmW}$, which is the ratio between the optical power in each sideband and the power of the optical carrier. $SBR$ depends on the experimental parameters as follows:
\begin{equation}
    {
    SBR = \frac{(\chi^{(2)} \cdot \omega_{o})^2\cdot \Gamma_{eo}^2\cdot L_{TL}^2\cdot PM^{2}}{2 \cdot n_{o}^2 \cdot n_{\Omega} \cdot c_0^3 \cdot \varepsilon_0 \cdot S_{\Omega}}  \cdot \eta P^{free-space}_{mmW}
    }
    \label{mod-eff-eq}
\end{equation}
where $\chi^{(2)}=\chi^{(2)}_{333}$, $\omega_{o}$ is the optical angular frequency, $n_{o}$ is the effective refractive index of the optical mode, $c_0$ is the speed of light in vacuum and $\varepsilon_0$ is the vacuum permittivity. 
$PM = \left| \frac{e^{i \Delta \Tilde{k} L_{TL} } - 1}{\Delta \Tilde{k} L_{TL}} \right| $ is the phase-matching function quantifying both mmWave loss and propagation phase mismatch acquired between the interacting fields along the transmission line's length $\Delta \Tilde{k} = \frac{2\pi f_{mmW}}{c_0} (n_{\Omega} - n_{g}) + i \frac{\alpha_{\Omega}}{2}$,  $\eta$ is the coupling efficiency of mmWave power into the transmission line being proportional to the square of $IAF$, and $P^{free-space}_{mmW}$ is the power of the mmWave beam in free-space. Using these definitions, the corresponding on-chip power is $P^{on-chip}_{mmW} = \eta \cdot P^{free-space}_{mmW}$ (complete derivation in Supplementary Information section \ref{SItheory}).

Our mmWave source provides up to 100 mW in the WR9.0 frequency range (82-125 GHz) and 5 mW in the WR2.8 frequency range (240-380 GHz), and this radiation is emitted through a horn antenna into free-space. MmWave signal is coupled into the chip using a backside-mounted silicon lens. The modulated optical signal is collected using a lensed fiber, and then visualized on an optical spectrum analyzer (Fig.~\ref{fig2}~a-b, details in methods).
Our wireless electro-optic modulators support efficient formation of sidebands across the entire incident mmWave range (Fig.~\ref{fig2}~c-d). To gain quantitative insight into the efficiency of our broadband modulators and eliminate the effect of the variation in mmWave power from the source, we analyze the frequency-dependent efficiency of the modulator by reporting the sideband ratio per unit of milliwatt incident mmWave power (Fig.~\ref{fig2}~e-f). We extract modulation efficiency values of ${10^{-4}}$/mW and ${3\cdot 10^{-5}}$/mW for configurations A and B, respectively. Using Eq.~\ref{mod-eff-eq}, we estimate the amount of mmWave power coupled from free-space to the chip to be 1-8\% across the WR9.0 band and 2-6\%  across the WR2.8 band, respectively (Fig.~\ref{fig2}~g-h). We associate the linear frequency dependence of the coupling efficiency with the frequency-dependent beam waist of the Gaussian mmWave beam generated by the horn antenna. We calculate a half-wave voltage-length product $V_\pi \cdot L_{TL} = \sqrt{2 Z_{TL} \frac{P^{on-chip}_{mmW}}{SBR}} \cdot L_{TL}$ of 3.4 V$\cdot$cm in WR9.0 and 3.65 V$\cdot$cm at WR2.8 band, comparable to previous studies using wired mmWave electro-optic modulators~\cite{wang2024ultrabroadband}.

\subsection*{Photonics-enabled characterization of mmWave transmission lines}
The performance of our wireless modulators strongly depends on their mmWave loss and the transmission line's refractive index. At microwave frequencies, these properties are routinely measured by performing a scattering matrix analysis using a vector network analyzer. For optimization purposes, it's essential to measure these values for our devices, but such instruments are expensive in the mmWaves, and their calibration requires a wired signal delivery. 

\begin{figure}[h!]
    \centering
    \includegraphics[width=17cm]{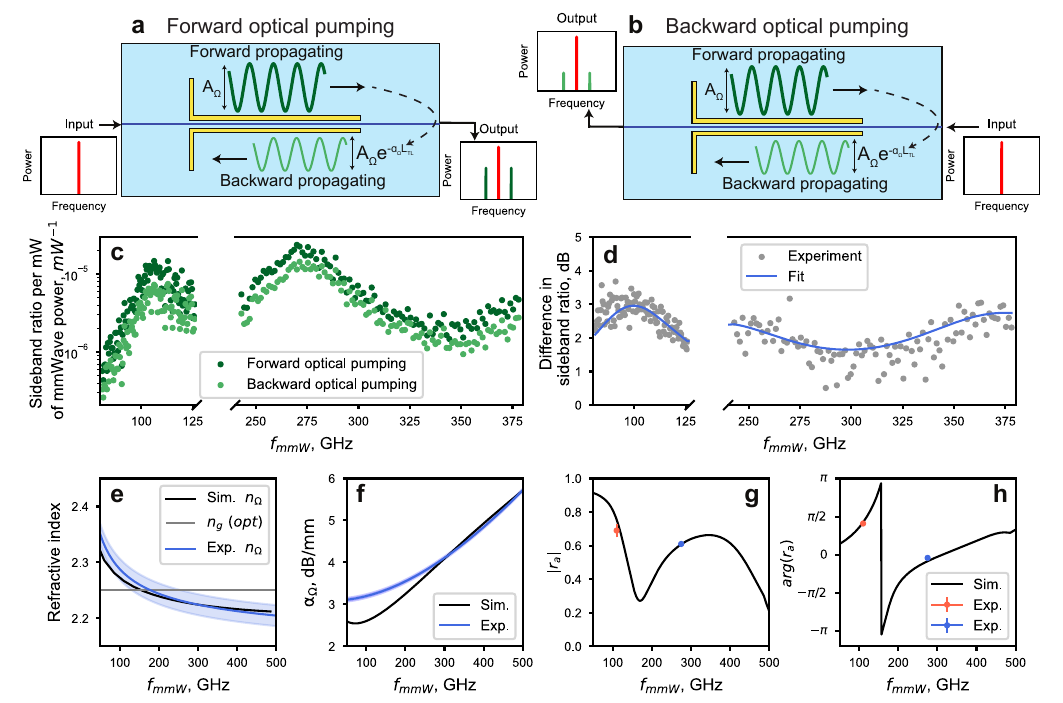}
    \renewcommand{\baselinestretch}{1} 
    \caption{\textbf{Photonics-enabled characterization of mmWave transmission lines.} (a) Schematic for the forward optical pumping scheme. The optical signal is coupled from the left, the mmWave signal couples through the antenna to the transmission line (amplitude $A_\Omega$) co-propagating with the optical signal in the forward direction. As the mmWave signal propagates along the transmission line, it gets attenuated due to loss $\alpha_{\Omega}$ while generating sidebands. By the end of the transmission line, the mmWave signal is reflected with approximately unity efficiency (amplitude $A_{\Omega}e^{-\alpha_\Omega L_{TL}}$). This counter-propagating mmWave signal also interacts with the optical signal with lower efficiency. Two processes (co- and counter-propagating optical and mmWave signals) lead to the formation of sidebands in the optical domain (dark green lines in the output spectra). (b) Sketch for the backward optical pumping scheme. In this case, the direction of propagation of the optical signal is swapped. Once the mmWave is reflected at the end of the transmission line, the optical and mmWave co-propagate. The optical signal now interacts with the attenuated mmWave signal, resulting in weaker sideband generation. (light green lines in the output spectra). We note that for long transmission lines ($L_{TL} > 0.5$ mm), only the co-propagating mmWave and optical signals are phase-matched. We employ this to extract the loss. (c) Normalized sideband ratio for the forward (dark green) and backward (light green) optical propagation. We observe a lower sideband ratio for the backward optical configuration, which allows us to estimate the transmission line loss $\alpha_{\Omega}$. (d) Ratio of forward to backward propagation sideband ratios (gray points), along with the fitted curve (blue curve). The clear frequency dependency facilitates the extraction of the refractive index of the mmWave mode propagating along the transmission line (e). Comparison between simulated and fitted values of the refractive indices. (f) Comparison between simulated and experimental values of the mmWave losses. (g) Absolute values of the antenna-transmission line reflection coefficient $r_a$ enabled by the frequency dependency of the sideband ratio. (h) Phase of the reflection coefficient.}
    \renewcommand{\baselinestretch}{1.5} 
    \label{fig3}   
\end{figure}

We propose a photonics-enabled technique to extract the transmission line's parameters. Our method involves performing two experiments, as shown in Fig.~\ref{fig3}~a and b. The direction of propagation of the optical signal is reversed between the two experiments, enabling interaction with the mmWave signal in either a forward optical or a backward optical configuration. In both cases, the mmWave travels forward along the transmission line, gets reflected at its end, and then returns. With phase-matching and sufficient single-pass mmWave loss (achieved by using long transmission lines), the mmWave signal is attenuated during its forward propagation, leading to an attenuation of the mmWave signal amplitude from $A_\Omega$ to $A_\Omega e^{-\alpha_\Omega L_{TL}}$. This makes the modulation more efficient in the forward optical configuration compared to the backward optical configuration, where the added attenuation of the mmWave signal during forward propagation leads to a smaller sideband ratio, see Fig.~\ref{fig3}~c for a transmission line length of $L_{TL}$ = 0.5 mm. In addition, for short transmission line lengths $L_{TL}$, phase-matching occurs for both co-propagating and counter-propagating mmWave and optical signals, and the total modulation is a sum of these two contributions. This interference introduces the sinusoidal modulation in the ratio of the sideband-ratio Fig.~\ref{fig3}~c. Relating the sideband ratios of these two measurements (shown in Fig.~\ref{fig3}~d) allows extracting the transmission line's refractive index, shown in Fig.~\ref{fig3}~e, and the transmission line's loss $\alpha_\Omega$, shown in Fig.~\ref{fig3}~f. The extracted values agree well with the simulated values. 

With experimental values for loss and refractive index now determined, a remaining parameter is the reflection coefficient at the antenna-transmission line interface. Given the impedance mismatch between the antenna and the transmission line, a weak cavity is formed by the transmission line, with a reflectivity dictated by the antenna on one end and the open circuit termination on the other end, leading to fringes in the sideband ratio in Fig.~\ref{fig3}~c. By comparing various $L_{TL}=$~0.25, 0.5 and 1~mm and exploiting the fact that reflection losses are unaffected by $L_{TL}$, whereas propagation loss increases linearly with $L_{TL}$, we extract the complex reflection coefficient of the antenna $r_a$. We find a reflection coefficient $\approx 0.69$ and $\approx 0.61$ around frequencies of 110 and 270 GHz, respectively (see Fig.~\ref{fig3}~g and h). Using these deduced parameters, we can model the experimental data shown in Fig.~\ref{fig2}~e-f by applying eq.~\ref {mod-eff-eq}, which shows good agreement (more details are provided in the Methods and Supplementary Information section \ref{ReflmodelSI}).

 We note that the mmWave loss in our devices can be reduced by increasing the thickness of the transmission line electrodes (see Supplementary Information section \ref{SIlossmodel}). Additionally, the overall efficiency of the system can be further improved by enhancing the coupling efficiency of the mmWave signals to the transmission line through the implementation of broadband antennas (see Supplementary Information section \ref{CouplEffStrategiesSI}).

\subsection*{High-speed mmWave detection}\label{fast_thz_detector}
    
In the context of a 6G communication scheme, a mmWave signal serves as the carrier, with data packets encoded onto it via amplitude or phase modulation. This modulation process broadens the signal's bandwidth. Specifically, it extends the carrier's frequency range by twice the modulation frequency. Consequently, detecting the full span of this modulated mmWave signal requires a broadband device capable of covering the entire resulting frequency spectrum. This bandwidth requirement also applies to other applications such as high-speed mmWave sensing.

\begin{figure}[h!]
    \centering
    \includegraphics[width=17cm]{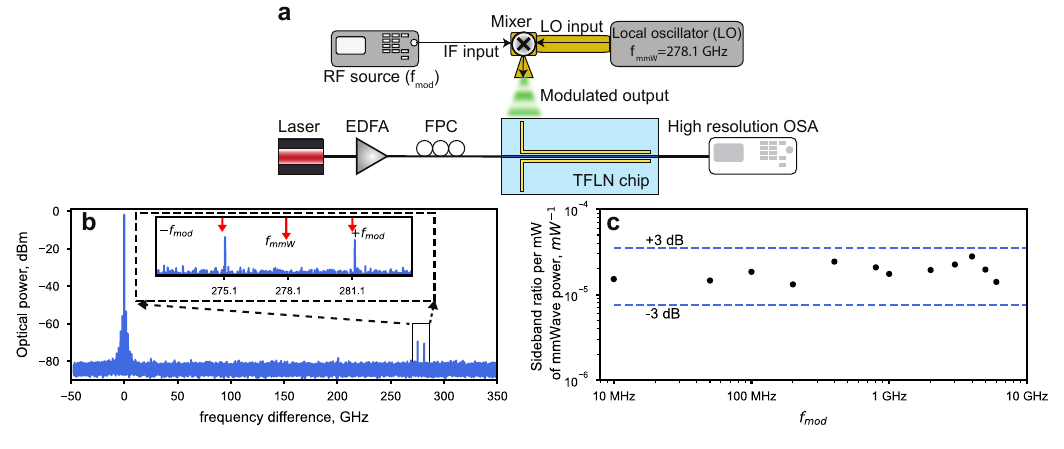}
    \renewcommand{\baselinestretch}{1} 
    \caption{\textbf{High-speed detection of wireless mmWave radiation} (a) Characterization setup for the traveling wave electro-optic modulator under illumination of an intensity-modulated mmWave signal (carrier frequency=278.1 GHz). The sinusoidal modulation ($f_{mod}$) and the mmWave signals are interfaced in a mixer to generate the modulated output. The mmWave is connected to the local oscillator (LO) port and the modulation signal is connected to the intermediate frequency (IF) port. The modulated mmWave signal illuminates the chip, which is pumped by an optical signal that passes through an erbium-doped fiber amplifier (EDFA) and a fiber polarization controller (FPC). The output modulated optical carrier is monitored using a high-resolution optical spectrum analyzer (OSA). (b) Optical spectrum at offset frequencies from the optical carrier, under the incident mmWave signal modulated at 3 GHz. Two optical tones offset by $f_{mmW}$ from the optical carrier and separated by $2\cdot f_{mod}$ confirm efficient intensity modulation of the mmWave and suppression of the $f_{mmW}$ tone. The inset shows the zoomed region, marked with a dashed black box. (c) Normalized sideband ratio as a function of modulation frequency, showing a flat modulation efficiency up to 6 GHz. Dashed lines represent $\pm$3~dB bounds.}
    \renewcommand{\baselinestretch}{1.5} 
    \label{fig4}  
\end{figure}

To verify the suitability of our modulator for these applications, we now implement this device as a high-speed broadband mmWave detector. In this configuration, we amplitude modulate a 278.1~GHz carrier using a high-speed electronic mixer. We note that this frequency has been chosen arbitrarily, and similar performance was observed for other carrier frequencies. To investigate the electronic bandwidth of our detector, we apply frequencies spanning from 10 MHz to 6 GHz to the mixer by connecting an RF source to the intermediate frequency (IF) port of the mixer. A sketch of the experimental setup is shown in Fig.~\ref{fig4}~a.

In this configuration, the generated mmWave power is -21~dBm (vs +5~dBm in the previous experiments), limited by the power handling of the electronic mixer (-11 dBm of IF power to maintain  <1 dB compression level) and its conversion loss (10 dB). Given the sideband ratio per mW of mmWave power (Fig.~\ref{fig2}~e), we expect a sideband ratio of approximately -67~dB (compared to -38~dB in the experiments discussed above). To overcome these challenging power constraints, the optical signal is preamplified using an Erbium-Doped Fiber amplifier (EDFA), resulting in an increased input optical power of +20~dBm before the chip, thereby increasing the sideband power. 

We investigate the bandwidth of our wireless phase modulators using configuration A (Fig. \ref{fig2} b) while monitoring the optical spectrum after the modulator on a high-resolution optical spectrum analyzer. For an amplitude-modulated mmWave signal, we observe two peaks at frequencies $f_{mmW}-f_{mod}$ and $f_{mmW}+f_{mod}$ (Fig.~\ref{fig4}~b). By sweeping the modulation frequency from 10 MHz to 6 GHz (currently limited by the experimental setup), we observe that the separation between the two peaks scales linearly with $f_{mod}$. The large modulation bandwidth of our modulators enables the amplitude of the sideband to remain relatively flat throughout the sweep, making it suitable for applications such as 6G or high-speed mmWave/terahertz detection (Fig.~\ref{fig4}~c).

\subsection*{MmWave electro-optic frequency combs by triply resonant nonlinear interaction}

Apart from traveling-wave electro-optic modulators, cavity-based modulators play a pivotal role in advancing mmWave photonics. By enabling optical photons to make multiple round-trips within the cavity, these modulators significantly enhance the photon-photon interaction probability compared to single-pass configurations. In addition, our findings on the antenna-transmission line reflection coefficient enable us to design a mmWave cavity, where the external coupling rate can be engineered through the antenna design, and the electro-optic interaction is further enhanced by multiple round-trips of mmWave photons within the cavity. In particular, cavity-based modulation schemes enable applications such as electro-optic transduction~\cite{holzgrafe2020cavity, mckenna2020cryogenic,  doi:10.1126/sciadv.aar4994} and ultra-sensitive field sensing~\cite{ma_integrated_2024} even under weak mmWave illumination. Under strong mmWave illumination, the generated photons can interact multiple times with the mmWave photons, leading to cascaded sideband generation and resulting in the formation of an optical frequency comb~\cite{rueda2019resonant, zhuang2023electro, zhang2025ultrabroadband}. 

\begin{figure}[h!]
    \centering
    \includegraphics[width=17cm]{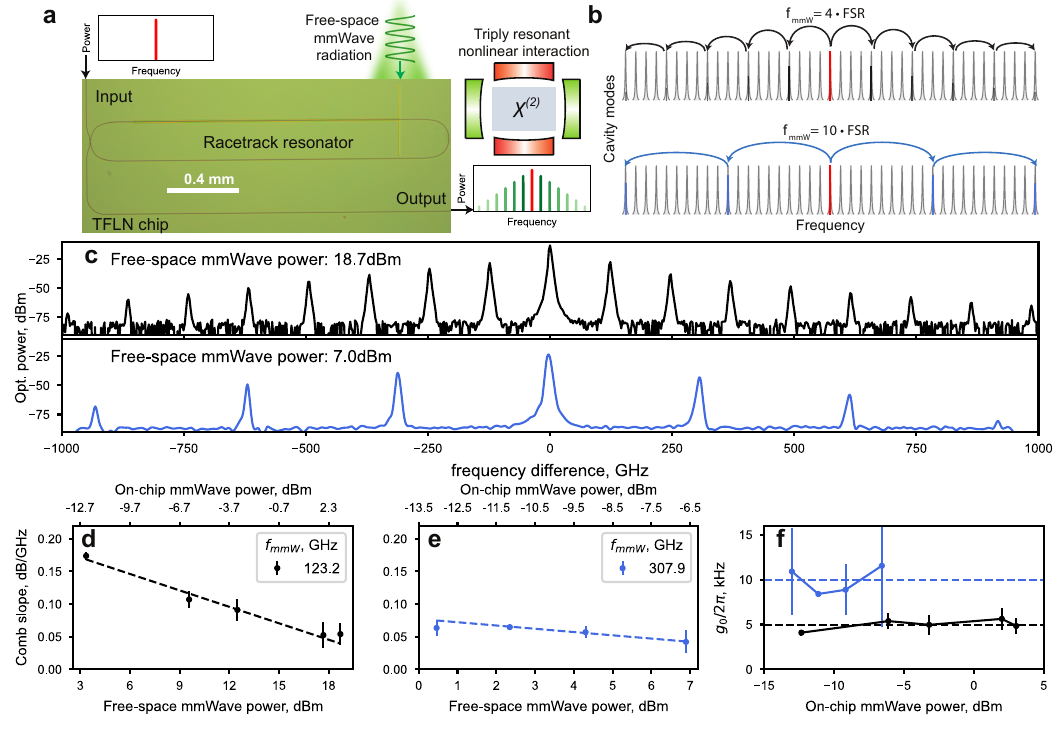}
    \renewcommand{\baselinestretch}{1} 
    \caption{\textbf{Wireless electro-optic frequency comb generation via triply resonant nonlinear interaction}. (a) A sketch of the setup. The continuous-wave laser pumps the racetrack resonator, and the laser's frequency matches one of the racetrack cavity's resonances. A mmWave phase modulator is patterned along one of the straight sections of the racetrack. Free-space mmWave radiation is incident on the chip, leading to phase modulation of optical photons passing multiple times through the transmission line in the cavity. This process cascades, and the generated sideband interacts with the mmWave photons, generating more sidebands, leading to the formation of an electro-optical comb spectrum. (b) A representation of the cascaded electro-optic comb generation in this work. The pumping laser (red line) pumps one of the optical resonances once the mmWave frequency $f_{mmW}$ matches an integer multiple of the racetrack's FSR (in this case multiples of 4, and 10), the generation process cascades leading to a larger electro-optic bandwidth. (c) The measured output spectra out of the chip upon mmWave radiation incidence at frequencies of 123.2 GHz = 4 $\cdot$ FSR (upper plot, black line) and 307.9 GHz = 10 $\cdot$ FSR (bottom plot, blue line) at maximal mmWave free-space powers; the comb slopes as the function of mmWave power at (d) 123.2 GHz and (e) 307.9 GHz; the upper x-axis indicating the calculated mmWave on-chip power; (f) calculated single-photon coupling rate $g_0$ at 307.9 GHz (blue curve) and 123.2 GHz (black curve); the dashed lines show the mean value over the whole power range.}
    \renewcommand{\baselinestretch}{1.5} 
    \label{fig5}  
\end{figure}

Resonant electro-optic frequency comb sources can be driven by multiples of the optical cavity's free spectral range (FSR). In practice, this is not straightforward to achieve with conventional electronics due to technical challenges in realizing modulators that operate up to hundreds of gigahertz bandwidth. However, our wideband modulator provides access to a broad range of frequencies.
To achieve a mmWave electro-optic frequency comb, we fabricate a racetrack resonator with a designed free-spectral range of the cavity of 30.79 GHz. We place a 1.5 mm-long mmWave antenna-coupled transmission line in one of the arms of the racetrack to phase modulate the light circulating in the cavity (Fig.~\ref{fig5}~a). We choose the length of the mmWave cavity to ensure the largest modulation efficiency, while also matching the mmWave cavity's resonant frequencies to multiples of the optical FSR (see details in Supplementary Information \ref{LossRatesSI}). We use configuration A to characterize our mmWave electro-optic comb source. Given that
\begin{itemize}
    \item the frequency of the optical radiation matches the resonant frequency of the cavity,
    \item the frequency of the mmWave radiation matches  $n \times FSR$ , where n is an integer number and FSR is free spectral range of the optical cavity, and
    \item the frequency of the mmWave matches the resonant frequency of the mmWave cavity,
\end{itemize}
The theoretical model shows that in the weak-pump regime, where electro-optic coupling rate ($g_{eo}$) is smaller than optical resonance linewidth ($\kappa$), the comb slope $S$ [dB/GHz] is given by~\cite{zhang2025ultrabroadband}:
\begin{equation}
\label{slope_eq}
    {
    S \approx -\frac{10}{f_{mmW}} \log_{10} \left( \frac{4 g_{eo}^2}{\kappa^2} \right).
    }
\end{equation}
Here, $g_{eo}=g_0 \sqrt{n_{mmW}}$, with $g_0$ being the single-photon electro-optic coupling rate and $n_{mmW}$ being the number of mmWave photons in the transmission line.

We design the microring resonator to be under-coupled and measure its linewidth to be $\kappa/2\pi = 220$~MHz (see Supplementary Information Section \ref{LwSI}). We record the output spectra of the racetrack resonator under two distinct illumination frequencies: 123.2~GHz, corresponding to 4×FSR, and 307.9~GHz, corresponding to 10×FSR of the ring (Fig.~\ref{fig5}b). The comb offers a characteristic slope of 0.05 dB/GHz (Fig.~\ref{fig5}~c) and spans a bandwidth of approximately 2~THz. We further verify that the comb slope scales with the incident free-space mmWave power at both illumination frequencies, showing a linear dependence consistent with Eq.~\ref{slope_eq} for both mmWave frequencies of 123.2 GHz and 307.9 GHz (Fig.~\ref{fig5}~d-e), see details in Supplementary Information section \ref{CouplEffRingsSI}. However, we find that achieving the same slope requires a lower power at 307.9~GHz than at 123.2~GHz (e.g. 2.7~dBm at 123.2~GHz versus -6.6~dBm at 307.9~GHz),  indicating that the efficiency of the comb generation per mW of mmWave power is larger at higher frequencies. We attribute this higher efficiency to the increased single-photon electro-optic coupling rate at higher frequencies, which scales as $g_0 \propto \sqrt{f_{mmW}}$ ~\cite{zhang2025ultrabroadband}. The estimated single-photon electro-optic coupling rates $g_0/2\pi$ from the experimental data are 4.98 $\pm$ 0.53 kHz and 9.93 $\pm$ 1.33 kHz at 123.2 and 307.9 GHz, respectively (Fig.~\ref{fig5}~f). These values are approximately 5 times higher than the previous reports~\cite{rueda2019resonant, multani2025integrated, zhang2025ultrabroadband} (see full analysis in Methods and Supplementary Information \ref{LossRatesSI}). We also extract $g_0$ from resonance splitting at 123.2 GHz, resulting in the same value of 4.79 $\pm$ 0.52 kHz (see details in Supplementary Information section \ref{ResonanceSplitting}).

\tocless\section{Conclusions and outlooks}\label{sec_conclusions}

In summary, we present a theoretical and experimental demonstration of a wireless electro-optic modulation scheme on the thin-film lithium niobate platform. In this approach, the mmWave radiation is coupled directly to the chip via an integrated antenna. This wireless coupling approach offers practical advantages, eliminating the need for impedance-matched transmission lines and avoiding the complexity and cost associated with high-frequency mmWave probes. We developed experimental design guidelines for implementing wireless modulators across broad mmWave frequency bands and introduced a photonics-based metrology technique to characterize key performance metrics, including the refractive index, transmission line loss, and the complex reflection coefficient at the antenna–transmission line interface. We demonstrated coupling efficiencies of up to 8\% and showcased a path towards further improvement through antenna engineering. In addition, we demonstrated a reflection coefficient of $\approx 0.69$ at and $\approx 0.61$ at frequencies around 100 and 270 GHz, respectively. The demonstrated reflection coefficients enable the implementation of mmWave cavities with loaded Q-factors of approximately 6 and 12 at 123~GHz and 308~GHz, respectively, corresponding to intrinsic Q-factors of about 8 and 16. Combining lithium niobate properties and strong mmWave confinement enabled a nearly fivefold enhancement in the single-photon electro-optic coupling rate $g_0$, compared to previously reported values.  

This novel architecture holds promise for broadening the scope and applicability of electro-optic modulators across a wide range of disciplines. In particular, this approach is ideally suited for mmWave and terahertz sensing applications, where compact, high-bandwidth, and low-loss modulators are essential for detecting weak electromagnetic fields. In addition, this platform may hold some potential for quantum technologies. The ability to perform high-frequency electro-optic modulation without direct wiring could become relevant for emerging qubit platforms ~\cite{anferov2024millimeter}. In such systems, minimizing thermal load, wiring density, and electromagnetic interference is critical, and the proposed wireless modulator offers a scalable path toward compact, low-noise quantum interfaces. However, looking ahead, increasing the intrinsic Q-factors of our on-chip mmWave/terahertz cavities is of great importance for quantum optics and quantum computing applications. Ways to achieve this would comprise increasing the thickness of the metallic electrodes, which, as discussed in Supplementary Information \ref{SIlossmodel}, would reduce conductive losses by a factor of three and potentially raise the intrinsic Q-factors to around 50. Moreover, operating the device at cryogenic temperatures with superconducting metals such as niobium could further suppress conductive losses, offering a pathway to significantly higher Q-factors.
Our results lay the foundation for compact, efficient, and scalable wireless electro-optic modulators, unlocking new possibilities in high-frequency sensing, free-space communication, and quantum information science.

\tocless\section{Methods}
\subsection*{Fabrication}
The chips are fabricated on 600 nm of X-cut lithium niobate, bonded to 4700 nm of thermally grown oxide on an approximately 500 $\mathrm{\mu m}$-thick double-side polished high-resistivity silicon substrate. The waveguides are patterned using electron-beam lithography (Eliionix ELS-HS50) and Ma-N resist. These waveguides are then etched into the LN layer using  $\mathrm{Ar^+}$ ions, followed by annealing in an $\mathrm{O_2}$ environment to recover implantation damages and improve the absorption loss of the platform \cite{shams2022reduced}. Subsequently, we clad the devices with 800 nm of Inductively Coupled Plasma - Chemical Vapor Deposition (Oxford Cobra), followed by another annealing. The electrodes are defined using a self-aligned process, including patterning with PMMA resist, dry etching, and lift-off using the same resist. The electrodes are deposited using electron-beam evaporation (Denton) with 15 nm of Ti and 300 nm of Au.

\subsection*{Experimental setup and measurements}

The tunable Keysight N7776C laser generates a 1550 nm continuous wave 16 mW pump, which is then coupled by lensed fiber (OZ Optics TSMJ-X-1550-9/125-0.25-7-2.5-14-2) to the lithium niobate waveguide on the chip from the edge. The polarization is aligned to the TE mode of lithium niobate waveguide by a Thorlabs FPC561 Fiber Polarization Controller. In the measurements (section \ref{fast_thz_detector}) of wireless electro-optic modulator bandwidth of the main text, there is also an EDFA (Erbium-Doped Fiber Amplifier) Nuphoton Technologies CW-C0-MR-30-20-FCA with a gain in power of 10. For the outcoupling we use (OZ Optics TSMJ-X-1550-9/125-0.25-7-2.5-28-5).

The mmWave radiation is generated the following way: an Anritsu RF/Microwave Signal Generator
MG362x1A sends an RF signal (9-14 GHz) to the mmWave source (Virginia Diodes Signal Generator Extension module, WR9.0 + WR2.8), which provides a 9x or 27x frequency up-conversion and emits up to 100 mW in WR9.0 range and up to 5 mW in WR2.8 range. The mmWave radiation is emitted into the free space via a horn antenna (Virginia Diodes WR8.0CH antenna is used in the frequency range of 80-125 GHz with a beam waist at central frequency: \~5.2~mm and WR2.8DH antenna in the frequency range of 240-380 GHz, beam waist at central frequency: \~1.9~mm). In the configuration A of the text we use 1 inch aperture, 10~mm focal length TPX lens to tightly focus the mmWave beam on the silicon lens, while in the configuration B we used a pair of TPX lenses: the first one, with an aperture of 1 inch and focal length of 20mm, is used to form a collimated beam after the output of the horn antenna, and the second one, with a large 2 inch aperture and 65mm focal length is used to focus the beam on the silicon lens. Then mmWave radiation is collected by the on-chip antenna and coupled into the transmission line. Thus, the optical and mmWave modes co-propagate and mix, creating sidebands detected by the OSA 
(Anritsu Optical Spectrum Analyzer MS9740B). For the measurements in section \ref{fast_thz_detector} the high-resolution OSA (Apex AP2043B) was used. The Keysight MXG Vector Signal Generator was used to generate the modulation microwave signal for the experiment in section \ref{fast_thz_detector}. A balanced mixer from Virginia Diodes (WR2.8BAMULP) was used to generate the modulated mmWave signal.
The TPX lens from Batop (LTA-D25.4-F10) was used to focus free-space mmWave radiation on the chip. However, in section \ref{subsec_antennamodelling}, for the measurements in configuration B, two lenses were used: first to collimate the mmWave beam (LTA-D25.4-F25), and then the second to focus the radiation on the chip (LTA-D50-F65). The Hyperhemispheric Silicon Lens from Batop (LSH-D12-T7.13) was fixed on the back of the chip using a self-designed chip holder. 

\subsection*{Photonics-enabled characterisation of mmWave transmission lines}

We implement a photonics-enabled technique to extract the transmission line's parameters. Our method relies on analyzing the difference in sideband ratio $dSBR$ is given by the following formula:
\begin{equation}
    dSBR = 10 \log_{10} \left( \frac{SBR_{forward}}{SBR_{backward}}\right) = 20 \cdot \log_{10} \left( 
    \frac{ 
    \left| 
    \frac{e^{i \Delta \Tilde{k}_{+} L_{TL}} - 1}{\Delta \Tilde{k}_{+} L_{TL}} 
    + 
    e^{i 2 k_\Omega L_{TL} - \alpha_\Omega L_{TL}} 
    \cdot 
    \frac{e^{i \Delta \Tilde{k}_{-} L_{TL}} - 1}{\Delta \Tilde{k}_{-} L_{TL}} 
    \right| 
    }
    { 
    \left| 
    \frac{e^{-i \Delta \Tilde{k}_{-} L_{TL}} - 1}{\Delta \Tilde{k}_{-} L_{TL}} 
    + 
    e^{i 2 k_\Omega L_{TL} - \alpha_\Omega L_{TL}} 
    \cdot 
    \frac{e^{-i \Delta \Tilde{k}_{+} L_{TL}} - 1}{\Delta \Tilde{k}_{+} L_{TL}} 
    \right| 
    }
    \right)
\end{equation}
which depends on the phase matching in co-propagating and counter propagating configuration, respectively, expressed through $\Delta \Tilde{k}_{\pm} = \frac{2\pi f_{mmW}}{c_0} (n_{\Omega} \mp n_{g}) \pm i \frac{\alpha_{\Omega}}{2}$. These two terms depend, in turn, on the terahertz refractive index $n_\Omega = n_0 + \frac{n'}{\sqrt{f_{mmW}}}$ where $n_0$ and $n'$ are the fitting parameters, and on the mmWave loss $\alpha_\Omega$. As we show in Supplementary Information \ref{SIlossmodel}, the main origin of the loss is the metallic loss caused by the finite surface impedance of the gold electrodes. Therefore, $\alpha_\Omega$ depends on $f_{mmW}$ and $\sigma_{Au}$ and geometrical parameters of the transmission line according to~\cite{hasnain1986dispersion, gallagher1987subpicosecond, keil1992high, martin2015artificial}. We note that the second term of the expression for the mmWave refractive index originates from the complex part of the surface impedance, as was explored in~\cite{keil1992high, cheng1994terahertz, phatak1990dispersion}. $n_g$ was kept to be 2.25 as measured in~\cite{lampert2024photonics}. Fitting this formula to our experimental resuls helps retrieve the mmWave refractive index and propagation loss of the transmission line. Our fitted value of gold conductance $\sigma^{fit}_{Au}=1.91\cdot10^7$ S/m, notably lower than the bulk value ($4.1\cdot10^7$ S/m), consistent with prior observations of evaporated thin films~\cite{lee_enhanced_2019}. 

We extract the reflection coefficient at the antenna-transmission line interface as follows. As derived in the Supplementary Information \ref{SItheory}, an impedance mismatch between the antenna and the transmission line will result in the mmWave signal being reflected at the antenna-transmission line interface, resulting in a weak cavity effect.  The transmission line forms a resonator with a reflectivity dictated by the antenna on one end and the open circuit termination on the other end. The resonator acts as a filter and only mmWave fields oscillating at frequencies that match its cavity modes can efficiently couple to the resonator, leading in visible fringes in the sideband ratio. The visibility of these fringes correlates with the reflection coefficient at the antenna-transmission line interface. Consequently, comparing various $L_{TL}$ allows extracting the complex reflection coefficient of the antenna $r_a$ which affects the fringe visibility, since reflection losses are unaffected by $L_{TL}$, whereas propagation losses increase linearly with $L_{TL}$. Then, the reflection coefficient $r_a$ is extracted by fitting the ratio of sideband ratios $SBR$ for devices of various lengths. More details are presented in the Supplementary Information \ref{ReflmodelSI}.

\subsection*{Calculation of the single-photon coupling rate $g_0$ from the electro-optic comb slope}\label{g0_calc}
First, the electro-optic coupling rate $g_{eo}$ is computed from the slope according to eq. \ref{slope_eq}. Then, the number of mmWave photons $n_{mmW}$ is estimated as follows~\cite{zhang2025ultrabroadband}:
\begin{equation}
    n_{mmW} = \frac{\kappa_{mmW, ex}}{\Delta_{mmW}^2 + \kappa_{mmW}^2/4} \frac{P^{on-chip}_{mmW}}{h f_{mmW}}
\end{equation}
where $h$ is the Planck constant, $\kappa_{mmW, ex}$ and $\kappa_{mmW}$ are external and full decay rates of the mmWave-cavity formed by a 1.5mm-long antenna-coupled transmission line, and $\Delta_{mmW}$ is the detuning of the mmWave frequency from the resonant frequency (details are presented in Supplementary Information \ref{LossRatesSI}).
And then, $g_0$ is estimated as $g_{eo}/\sqrt{n_{mmW}}$ and plotted in Fig.\ref{fig5}f. The errorbars for $g_0$ are calculated from the measurement error of the comb slope $S$.

\vspace{1cm}

\textbf{Data Availability}
The data generated in this study will be made available in the Zenodo database prior to publication.

\medskip
\textbf{Code Availability}
The code used to plot the data within this paper will be made available in the Zenodo database prior to publication.

\bibliographystyle{paper}
\bibliography{bibliography-main}

\medskip
\textbf{Acknowledgments}
A.G., Y.L., and I.C.B.C. acknowledge funding from the European Union’s Horizon Europe research and innovation programme under project MIRAQLS with grant agreement No. 101070700 and funding from the Swiss National Science Foundation Grant No. №219406. S.R. acknowledges funding from the Hans Eggenberger Foundation (independent research grant 2022, Switzerland) and the Swiss National Science Foundation (Postdoc Mobility, grant number 214483). L.M. acknowledges funding from the Behring Foundation and CAPES-Fulbright.

\medskip
\textbf{Author contributions} A.G. and I.C.B.C.  conceptualized the project. A.G. and K.M. built the optical setup, carried out the measurements, and performed the CST simulations. A.G. derived the theoretical description of the modulation efficiency. Y.L. designed the layout of the photonic chip. J.L. helped with the measurements of the electro-optic combs. A.S.-A., S.R., and L.M. fabricated the devices. A.G., A.S.-A. and I.C.B.C wrote the manuscript with feedback from all authors. I.C.B.C.. and M.L. supervised this work.

\medskip
\textbf{Competing interests}
The authors declare no competing interests. 

\medskip
\textbf{Disclaimer}
The views, opinions and/or findings
expressed are those of the author and should not be interpreted as representing the official views or policies of the Department of Defense or the U.S. Government.

\textbf{Corresponding authors} Correspondence to Aleksei Gaier (aleksei.gaier@epfl.ch) or Ileana-Cristina Benea-Chelmus (cristina.benea@epfl.ch). 

\appendix
\renewcommand{\thefigure}{S\arabic{figure}}
\setcounter{figure}{0} 
\setcounter{linenumber}{1}
\setcounter{equation}{0}
\renewcommand{\theequation}{S\arabic{equation}}
\clearpage 
\vspace*{1cm}

\begin{center}
    {\Huge \bfseries Supplementary Information}
\end{center}

\vspace{1cm}
\tableofcontents
\section{Chip details} \label{SIstructure}

\begin{figure}[h!]
    \centering
\includegraphics[width=0.8\linewidth]{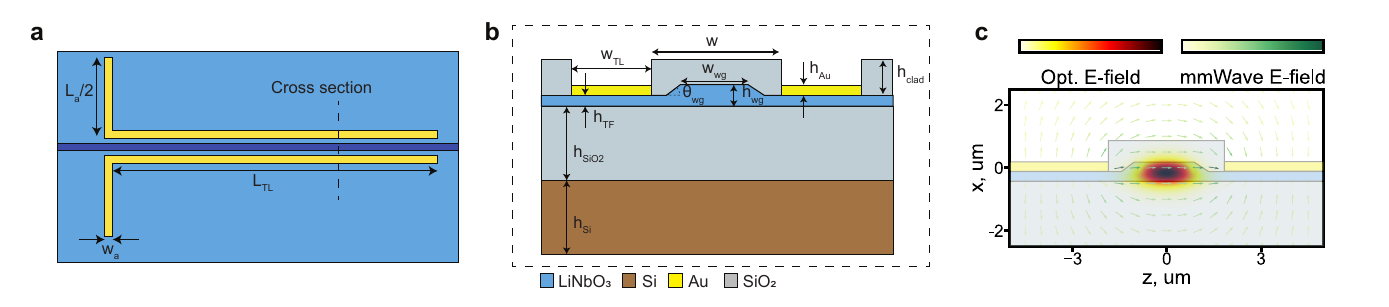}
    \caption{ (a) Top view of the dipole antenna devices showing the
transmission line with length $L_{TL}$. (b) Cross-section of the fabricated chips. (c) Simulated 2-D  profiles of the optical and mmWave modes.
}
    \label{fig_sample}
\end{figure}

Details on the geometry of the measured and simulated devices are given in Fig. \ref{fig_sample}, and all dimensions are provided in Table \ref{tab:dimensions}. The refractive index of lithium niobate, silicon, and silicon oxide is calculated using the Lorentz model from~\cite{wu2015temperature, naftaly2007terahertz}. 

\begin{table} [h!]
    \centering
    \begin{tabular}{ccc}
         thickness of the high resistivity silicon substrate & $h_{Si}$ & 500 ${\mu }$m\\ 
         thickness of the silicon oxide layer & $h_{SiO_2}$ & 4.7 ${\mu }$m \\
         thickness of thin-film lithium niobate layer & $h_{TF}$  & 300 nm \\
         thickness of the gold layer & $h_{Au}$  & 300 nm \\
         height of the lithium niobate waveguide & $h_{wg}$  & 600 nm \\
         thickness of the silicon oxide cladding & $h_{clad}$  & 1 $\mu $m \\
         width of the lithium niobate waveguide & $w_{wg}$  & 1.5 $\mu $m \\
         etching angle of the lithium niobate waveguide & $\theta_{wg}$ & $\ang{60}$
         \\
         distance between the electrodes & $w$  & 3.3 ${\mu }$m \\
         width of the antenna & $w_{a}$  & 5 $\mu $m \\
         width of the gold layer inside the transmission line & $w_{TL}$  & 3.5 $\mu $m \\
         length of the transmission line & $L_{TL}$  & 0.25-2 mm \\
         length of the dipole antenna & $L_{a}$  & 400 $\mu $m \\
    \end{tabular}
    \caption{Dimensions of the simulated and fabricated devices}
    \label{tab:dimensions}
\end{table}

\section{Theoretical model}
\label{SItheory}

In this chapter, we developed a theoretical description of the interaction of mmWave radiation coupled to the transmission line. First, we start with a general description of the nonlinear interaction between optical and mmWave modes. Then, we derive an expression for the amplitude of the mmWave E-field coupled to the transmission line via an on-chip antenna.

\subsection{Nonlinear frequency conversion mmWave-optics}
Due to the nonlinear mixing of the optical mode (at frequency $\omega_p$) with the mmWave mode (at frequency $\Omega$), the sidebands at frequencies $\omega_p \pm \Omega$ are generated.

We consider the linearly polarized optical fields along the z-axis (matching with the z-axis of the lithium niobate crystal), propagating along the y-axis. Then, the electric field might be written as follows:
\begin{equation}
    {E_{opt}(x,y,z,t) = \frac{1}{2} A(y) g_{opt}(x, z) e^{i(k y - \omega t)} + c.c.}
\end{equation}
where $A(y)$ is the electric field amplitude, $k=n_{o} \frac{\omega}{c_0}$ is the propagation constant of the optical eigenmode with the effective refractive index $n_{o}$ and the speed of light in vacuum $c_0$,  $g_{opt} (x,z)$ is a 2D field profile of the eigenmode of the waveguide, normalized in the following way: $\iint |g_{opt}|^2(x,z) dxdz = S_{opt}$, where $S_{opt}$ is the effective mode area. 
We can assume that the nonlinear interaction is weak, which implies the nondepletion case of the optical probe, which means that the probe field $E_{p}$ and the sideband $E_{SB}$ could be described as follows:
\begin{equation}
    E_{p}(x,y,z,t) = \frac{1}{2} A_p g_{opt}(x, z) e^{i(k_p y - \omega_p t)} + c.c.
\end{equation}
We assume $A_p = const$ with respect to $y$, where the subscript "p" stands for the probe mode.
\begin{equation}
    {E_{SB}^{(\pm)}(x,y,z,t) = \frac{1}{2} A_{SB^{(\pm)}}(y) g_{opt}(x, z) e^{i(k_{SB^{(\pm)}} y - \omega_{SB^{(\pm)}} t)}  + c.c.}
\end{equation}
where $\omega_{SB^{(\pm)}} = \omega_p \pm \Omega$, subscript "SB" stands for sideband mode, supscript $(\pm)$ stands for the upconverted (+) and downconverted (-) sideband modes. The mmWave electric field (also linearly polarized along z-axis) can be described as follows:
\begin{equation}
    E_{\Omega}(x, y, z, t) = 
    \frac{1}{2} A^+_{\Omega} \, g_{\Omega}(x, z) \, 
    e^{i(k_{\Omega} y - \Omega t)} \, 
    e^{-\frac{\alpha_{\Omega}}{2} y}
    + 
    \frac{1}{2} A_{\Omega}^{-} \, g_{\Omega}(x, z) \, 
    e^{-i(k_{\Omega} y + \Omega t)} \, 
    e^{\frac{\alpha_{\Omega}}{2} y} 
    + \text{c.c.}
\end{equation}
where $k_{\Omega}=n_{\Omega} \frac{\Omega}{c_0}$ is the wave vector of the RF wave with an effective refractive index $n_{\Omega}$, $\alpha_{\Omega}$ is the propagation loss of the mmWave radiation, and $A^+_{\Omega}$ and $A_{\Omega}^{-}$ are the amplitudes of the mmWave waves that co- and counter-propagate with an optical waves. Here we neglect the depletion of the mmWave amplitude due to the nonlinear interaction. The 2D profile of the mmWave mode is given by $g_{\Omega}$ function, which is normalized in the same way as $g_{opt}$: $\iint |g_{\Omega}|^2(x,z) dxdz = S_{\Omega}$.

The evolution of the optical electric field $E=E_p+E_{SB^+}+E_{SB^-}$ is given by the nonlinear wave equation:
\begin{equation} \label{nlwpeq}
    {\left( \nabla^2 - \frac{1}{c_0^2} \frac{\partial^2}{\partial t^2 }\right) E(x,y,z,t) = \frac{1}{\varepsilon_0 c_0^2} \frac{\partial^2}{\partial t^2 } \left( \mathcal{P}^L(x,y,z,t) + \mathcal{P}^{NL}(x,y,z,t) \right) }
\end{equation}
where ${\mathcal{P}^L(x,y,z,t) = \varepsilon_0 \int_{-\infty}^t\chi^{(1)}(x,y,z,t-t')E(x,y,z,t')dt'}$ is the linear part of the polarization with ${\chi^{(1)}}$ being linear susceptibility, and ${\mathcal{P}^{NL}(x,y,z,t)  =  \varepsilon_0 \chi^{(2)}_{333}(x,y,z,t)\cdot \left( E(x,y,z,t)+E_{\Omega}(x,y,z,t) \right)^2}$ is the nonlinear part of the polarization with ${\chi^{(2)}_{333}}$ being second-order nonlinear susceptibility. To simplify the expression for linear polarization, assuming the homogeneity of the material properties along the y-axis and using the properties of the Fourier transform, one may obtain:
\begin{equation}
\begin{split}
    {\mathcal{P}^L(x,y,z,t) = \varepsilon_0 \int_{-\infty}^t\chi^{(1)}(x,z,t-t')E(x,y,z,t')dt' = } \\ 
    {\frac{1}{2} \varepsilon_0 \chi^{(1)}(x,z,\omega_p) A_p g_{opt}(x, z) \cdot e^{i(k_{p} y - \omega_p t)} + } \\ 
     {\frac{1}{2}\varepsilon_0 \chi^{(1)}(x,z,\omega_{SB^+}) A_{SB^+}(y) g_{opt}(x, z) \cdot e^{i(k_{SB^+} y - \omega_{SB^+} t)}  + } \\ 
     {\frac{1}{2}\varepsilon_0 \chi^{(1)}(x,z,\omega_{SB^-}) A_{SB^-}(y) g_{opt}(x, z) \cdot e^{i(k_{SB^-} y - \omega_{SB^-} t)} + c.c.}
\end{split}
\end{equation}
Computing the second derivatives and using orthogonality of the complex exponents ${\int e^{i (\omega - \omega') t} dt = 2 \pi \delta(\omega-\omega')}$, we can rewrite the equation \ref{nlwpeq} to describe the evolution of the (+) sideband:
\begin{equation} 
\begin{split}
    {\left( \nabla^2 + \frac{\omega_{SB^+}^{2}}{c_0^2} n^2(x,z,\omega_{SB^+}) \right) A_{SB^+}(y) \cdot g_{opt}(x, z) \cdot e^{ik_{SB^+} y} = } \\
    {-\frac{\chi^{(2)}_{333}(x,z) \omega_{SB^+}^{2} }{c_0^2} A_p g_{opt}(x,z) g_{\Omega}(x,z) \left(A^+_{\Omega} e^{-\frac{\alpha_{\Omega}}{2} y}  e^{i(k_p + k_{\Omega})y} + A_{\Omega}^{-} e^{\frac{\alpha_{\Omega} y}{2}} e^{i(k_p - k_{\Omega})y}  \right)
    }
\end{split}
\end{equation}
Now, we can split the nabla operator into longitudinal and transversal components: ${\nabla^2 = \Delta_{\perp}^2 + \frac{\partial^2}{\partial y^2}}$, so that:
\begin{equation}
    {\left( \nabla^2 + \frac{\omega_{SB^+}^{2}}{c_0^2} n^2(x,z,\omega_{SB^+}) \right) g_{opt} (x,z) = \left( \frac{\partial^2}{\partial y^2} + \left( \frac{\omega_{SB^+}}{c_0} n_{o} \right)^2 \right) g_{opt} (x,z)}
\end{equation}
Multiplying both sides of the equation by ${g_{opt}(x,z)}$ and integrating over the coordinates ${(x,z)}$, one may obtain the following:
\begin{equation}
    {\left( \frac{\partial^2}{\partial y^2} + \frac{\omega_{SB^+}^{2}}{c_0^2} n_{o}^2 \right) A_{SB^+}(y) \cdot e^{ik_{SB^+} y} = 
    -\frac{\chi^{(2)} \omega_{SB^+}^{2} }{c_0^2} A_p \Gamma_{eo} \left(A^+_{\Omega} e^{-\frac{\alpha_{\Omega}}{2} y}  e^{i(k_p + k_{\Omega})y} + A_{\Omega}^{-} e^{\frac{\alpha_{\Omega}}{2}y} e^{i(k_p - k_{\Omega})y}  \right)}
\end{equation}
where we introduced the spatial overlap factor as follows:
\begin{equation} \label{overlapfactor}
    {\Gamma_{eo} = \frac{\iint_{LN} g^2_{opt} (x,z) \cdot g_{\Omega} (x,z) dx dz}{\iint_{x, z} g^2_{opt} (x,z) dx dz}}    
\end{equation}
where the integral in the nominator is taken over the nonlinear medium (in our case, lithium niobate waveguide), and the area of the integral in the denominator is the whole XZ plane, and we introduced effective nonlinear susceptibility $\chi^{(2)}=\chi^{(2)}_{333}$. Calculating the second derivatives in the left part of the equation gives:
\begin{equation} 
    {\left( \frac{\partial^2}{\partial y^2} + \frac{\omega_{SB^+}^{2}}{c_0^2} n_{o}^2 \right) A_{SB^+}(y) \cdot e^{ik_{SB^+} y}  =}  {e^{ik_{SB^+} y} \cdot \left( \frac{\partial^2 A_{SB^+}}{\partial y^2} + 2 ik_{SB^+} \frac{\partial A_{SB^+}}{\partial y} \right)}
\end{equation}
Substituting this result into the nonlinear wave equation gives: 
\begin{equation}\label{propeq}
    {\frac{\partial^2 A_{SB^+}}{\partial y^2} + 2ik_{SB^+}\frac{\partial A_{SB^+}}{\partial y}  = 
    -\frac{\chi^{(2)} \omega_{SB^+}^{2} }{c_0^2} A_p \Gamma_{eo} \left(A^+_{\Omega} e^{i\Delta \Tilde{k}_{+} y} + A_{\Omega}^{-} e^{i\Delta \Tilde{k}_{-} y}  \right)}  
\end{equation}
where we introduced $\Delta \Tilde{k}_{\pm} = \frac{2\pi f_{mmW}}{c_0} (n_{\Omega} \mp n_{g}) \pm i \frac{\alpha_{\Omega}}{2}$, with the group index of the optical mode ${n_g}$. If now we apply the slowly varying envelope approximation assuming ${\lvert \frac{\partial^2 A_{SB^+}}{\partial y^2} \rvert \ll \lvert 2 k_{SB^+} \frac{\partial A_{SB^+}}{\partial y} \rvert}$ for eq. \ref{propeq}:
\begin{equation} \label{propeqSVEA}
    { \frac{\partial A^+_{SB^+}}{\partial y}  = 
    i\frac{\chi^{(2)} \omega_{SB^+} }{2c_0 n_{o}} A_p \Gamma_{eo} \cdot \left(A_{\Omega}^+  e^{i\Delta \Tilde{k}_{+} y} + A_{\Omega}^{-} e^{i\Delta \Tilde{k}_{-} y}  \right)}     
\end{equation}
The solution of this equation at coordinate $y=L_{TL}$, assuming that at y=0 the sideband mode is not populated, or in other words that ${A_{SB^+}(y=0)=0}$, is given by:
\begin{equation} \label{SolutionNotneglected}
    {A_{SB^+} = \frac{\chi^{(2)} \omega_{SB^+} }{2n_{o}c_0} A_p  \Gamma_{eo} L_{TL} \cdot  \left( A^+_{\Omega} \frac{e^{i \Delta \Tilde{k}_+ L_{TL} } - 1}{\Delta \Tilde{k}_+ L_{TL}} + A_{\Omega}^{-} \frac{e^{i \Delta \Tilde{k}_{-} L_{TL} } - 1}{\Delta \Tilde{k}_{-} L_{TL}}\right)}     
\end{equation}
Note that this solution is valid only for small modulation, i.e. ${|A_{SB^+}|\ll|A_p|}$. All parameters are known in this formula, besides the amplitude of the mmWave radiation coupled to the transmission line via an on-chip antenna. Our next step is to find the relation between $A_\Omega^{\pm}$ and the incident electric field strength $E_{inc}$.

\subsection{Inverse antenna factor}
\label{suppl:coupled_THz_fields}
\begin{figure}[h!]
    \centering
    \includegraphics[width=0.6\linewidth]{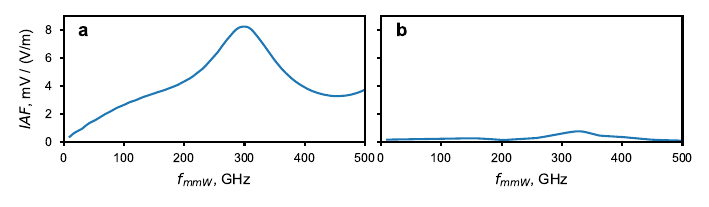}
    \caption{Simulated inverse antenna factors for the antenna design used in this work in two configurations: with (a) and without (b) Si lens.}
    \label{fig:IAF}
\end{figure}

Consider the incident electric field ${E_{inc}}$ being a plane wave. It induces a voltage between antenna pads $V_a$, which proportional to $E_{inc}$ by the inverse antenna factor ($IAF$):
\begin{equation}
    {
    V_{a} = IAF \cdot E_{inc}
    }
\end{equation}
This factor and its frequency dependence primarily depend on the metal conductivity and geometry of the antenna, and can be engineered to maximize the antenna's response in the desired frequency band.
The simulated values of the inverse antenna factor for the antenna used in this work are shown in Fig.~\ref{fig:IAF}. $IAF$ has a clear resonance at center frequency $\approx$300GHz, and reaching the value of 8 mV/(V/m). In this plot, we compare two ways of antenna illumination: with and without a silicon lens. Adding the silicon lens improved the antenna factor almost ten times. However, in practice, only if the input beam waist is much larger than the antenna aperture, this formula applies. In the case, for example, of antenna illumination with a Gaussian beam with a waist smaller than the silicon lens aperture, this value may differ from the simulated values and should be studied for the parameters of the input beam. 

The voltage generated by the antenna launches a voltage wave into the transmission line, and the relation between antenna voltage and the transmission line is studied below.

\subsection{Equivalent circuit of the antenna-coupled transmission line: relating voltages to electric fields}
\label{suppl:coupled_THz_fields_beginning}
\begin{figure}[h!]
    \centering
    \includegraphics[width=0.8\linewidth]{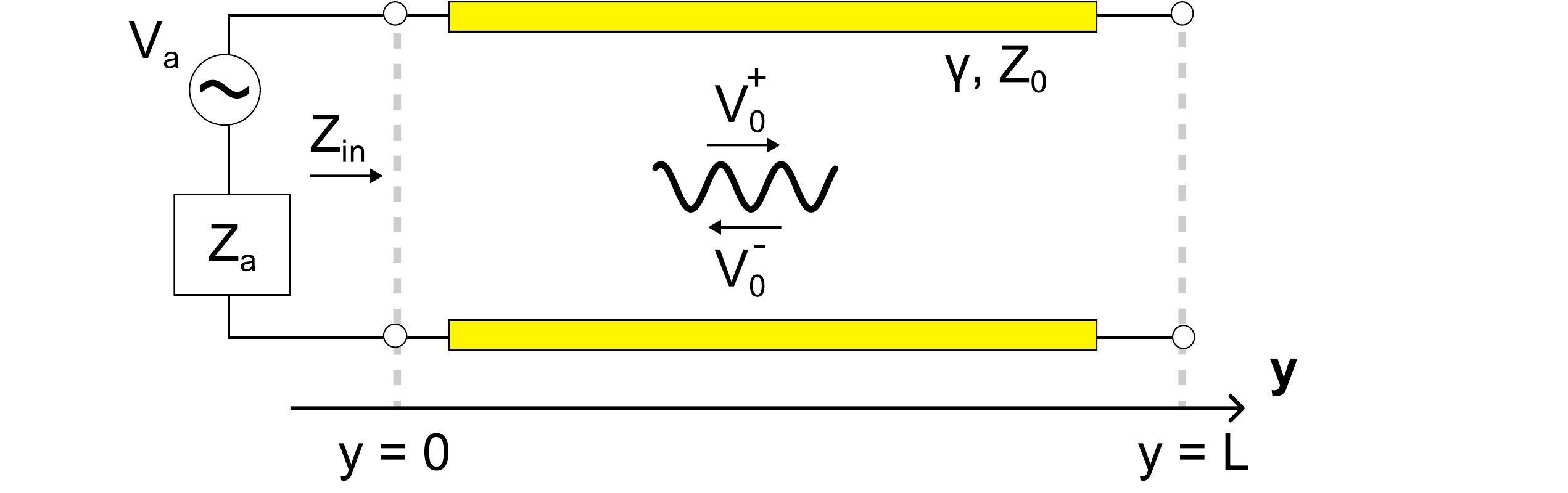}
    \caption{Equivalent circuit scheme for the devices under study. The antenna serves as a voltage generator connected in series to the transmission line. The reflection coefficient ${r}$ from the end of the transmission line is close to 1, thus an open circuit termination is schematically shown.}
    \label{fig:2}
\end{figure}

Consider the equivalent circuit for the devices with the antenna placed at the beginning of the transmission line shown in Fig.~\ref{fig:2}, where the line is terminated with a reflection coefficient ${r}$ at ${y=L}$. The antenna serves as a voltage generator with ${V_{a}}$ and ${Z_{a}}$. The present circuit could be studied using an infinite series to represent the multiple reflections from both ends of the transmission line. We use the impedance transformation method described in \cite{Pozar:882338}. The voltage and current in the line are given by the superposition of the waves propagating in the positive and negative directions along the transmission line:
\begin{gather}\label{V_def}
{V(y) = V_0^{+}e^{-\gamma y} + V_0^{-}e^{\gamma y}},\\
{I(y) = \frac{1}{Z_{TL}}(V_0^{+}e^{-\gamma y} - V_0^{-}e^{\gamma y}),}
\end{gather}
The reflection coefficient ${r}$ at the load (${y=L}$) is the ratio of amplitudes of the left-traveling wave and the right-traveling one:
\begin{gather}\label{refl_def}
{r = \frac{V_0^{-}e^{\gamma L}}{V_0^{+}e^{-\gamma L}} = \frac{V_0^{-}}{V_0^{+}}e^{2\gamma L}}
\end{gather}
Putting this expression into the ones above gives:
\begin{gather}
{V(y) = V_0^{+}(e^{-\gamma y} + r e^{-2\gamma L}e^{\gamma y})}, \\
{I(y) = \frac{V_0^{+}}{Z_{TL}}(e^{-\gamma y} - r e^{-2\gamma L}e^{\gamma y})},\\
{Z_{in}(y=0) = \frac{V(0)}{I(0)} = Z_{TL}\frac{1 + r e^{-2\gamma L}}{1 - re^{-2\gamma L}}},
\end{gather}
where ${\gamma = \frac{\alpha_{\Omega}}{2} - i k_\Omega} $ (${\alpha_{\Omega} > 0}$) is the complex propagation constant of the mmWave mode, ${Z_{TL}}$ is the characteristic impedance of the line.  

Since the phase-matching condition is satisfied primarily for the right-traveling wave, further we are interested in ${V_0^{+}}$. To find ${V_0^{+}}$ we note that on the one hand, ${V(0) =  V_0^{+}(1 + r e^{-2\gamma L})}$; on the other hand, from the antenna side we can write based on the expression for impedances connected in series: 
${V(0) = V_{a}\frac{Z_{in}}{Z_{in} + Z_{a}}}$. Equating these two expressions for ${V(0)}$ we get:
\begin{equation}\label{V0+_l=0}
{V_0^{+} = V_{a}\frac{Z_{in}}{Z_{in} + Z_{a}}\frac{1}{1+r e^{-2\gamma L}}}
\end{equation}
It's useful to rewrite eq.~\ref{V0+_l=0} in terms of reflection coefficients from both ends:
\begin{gather}
\label{V0+_l=0_simplified}
{V_0^{+} = V_{a}\frac{Z_{in}}{Z_{in} + Z_{a}}\frac{1}{1+r e^{-2\gamma L}} = V_{a}\frac{Z_{TL}\frac{1 + r e^{-2\gamma L}}{1 - r e^{-2\gamma L}}}{Z_{TL}\frac{1 + r e^{-2\gamma L}}{1 - r e^{-2\gamma L}} + Z_{a}}\frac{1}{1+r e^{-2\gamma L}} = V_{a}\frac{Z_{TL}\frac{1}{1 - r e^{-2\gamma L}}}{Z_{TL}\frac{1 + r e^{-2\gamma L}}{1 - r e^{-2\gamma L}} + Z_{a}} }= \\ \notag = { V_{a}\frac{Z_{TL}}{Z_{TL}(1 + r e^{-2\gamma L}) + Z_{a}(1 - r e^{-2\gamma L})} =V_{a}\frac{Z_{TL}}{Z_{TL} + Z_{a} + r e^{-2\gamma L}(Z_{TL} - Z_{a})} =} \\ \notag {= V_{a}\frac{Z_{TL}}{Z_{TL} + Z_{a}}\frac{1}{1 - r\cdot r_{a} e^{-2\gamma L}} = \frac{V_{a}}{2}  \frac{1-r_{a}}{1 - r_{a} \cdot a \cdot e^{i 2 k_{\Omega} L}}}
\end{gather}
where ${r_{a}}$ is the reflection coefficient from the antenna end of the transmission line: ${r_{a} = \frac{Z_{a} - Z_{TL}}{Z_{a} + Z_{TL}}}$, and $a=r\cdot e^{-\alpha_\Omega L}$ is a round-trip loss term. Thus,
\begin{gather}
\label{V0+_l=0_simplified}
{V_0^{+} = \frac{V_{a}}{2} \cdot \frac{1-r_{a}}{1 - r_{a} \cdot a \cdot e^{i 2 k_{\Omega} L}} = \frac{V_{a}}{2} \cdot t_a}
\end{gather}
where we introduced the transmission term (from the antenna to the transmission line) $t_a$. Using relation \ref{refl_def} we can obtain a formula for $V_0^-$:
\begin{gather}
\label{V0-_l=0_simplified}
{V_0^{-} = \frac{V_{a}}{2} \cdot t_a e^{-\alpha_\Omega L + i 2 k_{\Omega} L }}
\end{gather}
One can see that these formulas contain the sum of an infinite geometric series corresponding to the multiple reflections from both ends of the transmission line. The result represents a cavity where one mirror is the circuit termination, and the other is the antenna-transmission line interface. The simulated values of ${r}$ are close to unity, meaning that the end of the transmission line can be approximately considered an open circuit termination. To better understand the physical meaning of the formula, let us consider two extreme cases for the transmission term.

1. Let ${r_{a} = 1}$. This would happen if  ${Z_{a} \gg Z_{TL}}$. Then ${V_0^{+}=V_0^- = 0}$, meaning that all the mmWave field was reflected at the "antenna - transmission line" interface, and no voltage entered the transmission line.

2. Let ${Z_{a} = Z_{TL}}$, thus ${r_{a} = 0}$ and the transmission line is matched to the antenna. Then ${V_0^{+} = \frac{V_{a}}{2}}$ and ${V_0^{-} = \frac{V_{a}}{2}}e^{-\alpha_\Omega L + i 2 k_{\Omega} L }$.

\begin{figure}[h!]
    \centering
    \includegraphics[width=0.8\linewidth]{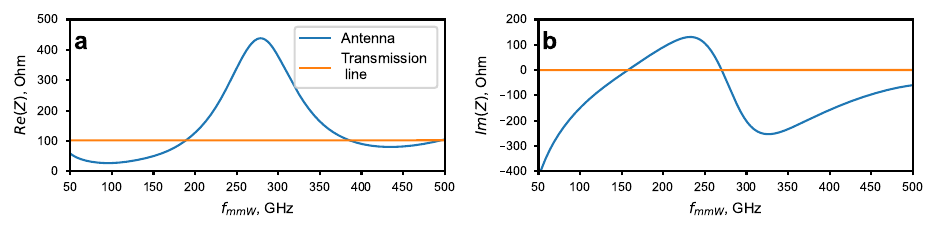}
    \caption{The simulated real (a) and imaginary parts (b) of the antenna and transmission line impedance.}
    \label{ReflCoefplot}
\end{figure}

We used the CST Studio software to simulate the impedances of the antenna used in this work and the transmission line. We plot them in Fig.\ref{ReflCoefplot}. Near the antenna resonance ($\approx 300$ GHz, where $IAF$ is maximized), the antenna has a relatively high real part of the impedance, resulting in a reflection coefficient $|r_a|\approx0.6$, while at frequencies below 150 GHz, the imaginary part of the impedance becomes dominant leading to an absolute value of the reflection coefficient close to 1.

The amplitudes of the electric field of mmWave traveling to the right and to the left ${A^{\pm}_{\Omega}}$ might be found from the two forms of the expression for the power of the mmWave mode ${P^{on-chip_{\pm}}_{mmW}}$:
\begin{equation}
    {
    {P^{on-chip_\pm}_{mmW}} = \frac{|V_0^{\pm}|^2}{2} Re \left( \frac{1}{Z_{TL}} \right) = \frac{1}{2} \varepsilon_0 n_{\Omega} c_0 |A_{\Omega}^\pm|^2 S_{\Omega}
    }
\end{equation}
Our simulations have shown the imaginary part of the characteristic impedance is negligible (Fig. \ref{ReflCoefplot}), so then ${Re  \left( \frac{1}{Z_{TL}} \right) \approx \frac{1}{Z_{TL}}}$. This gives the relation between the amplitude of the mmWave field in the transmission line ${A^\pm_{\Omega}}$ and the incident mmWave electric field ${E_{inc}}$:
\begin{gather}
\label{ATHz}
    {
    A^+_{\Omega} = \frac{IAF \cdot t_a}{2 \sqrt{Z_{TL} \varepsilon_0 n_{\Omega} c_0 S_{\Omega}}}  E_{inc}}, \\
    {A^-_{\Omega} = A^+_{\Omega}e^{-\alpha_\Omega L + i 2 k_{\Omega} L}
    }
\end{gather}

\begin{figure}[h!]
    \centering
    \includegraphics[width=0.8\linewidth]{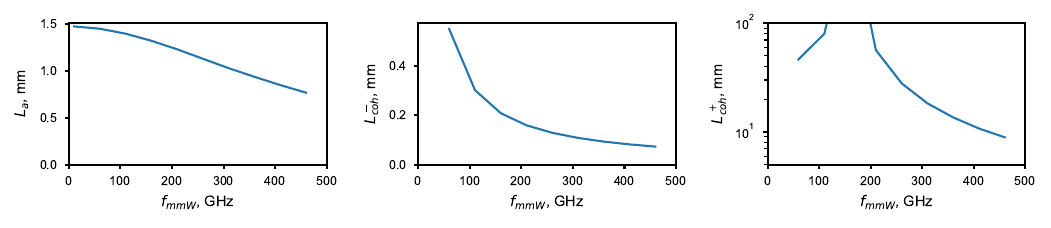}
    \caption{\textbf{Simulated coherence and absorption lengths.} (a) Absorption length $L_a = 1/\alpha_\Omega$, Coherence lengths (b) $L_{coh}^- = \frac{c_0}{2f_{mmW}(n_\Omega+n_g)}$ and (c) $L_{coh}^+ = \frac{c_0}{2f_{mmW}|n_\Omega-n_g|}$ }
    \label{SimLengths}
\end{figure}

These expressions clearly indicate that depending on the device's length $L$, term $t_a$ changes significantly. However, not only does the amplitude of the reflected wave depend on the device's length, but the phase-matching conditions for right- and left-traveling waves also differ, as indicated in eq.\ref{SolutionNotneglected}. In other words, there are two length scales: one is related to the amplitude of the left-traveling wave ($L_a = 1/\alpha_\Omega$), and the other is associated with the efficiency of interaction between right-traveling optical mode and left-traveling mmWave mode $L_{coh}^- = \frac{c_0}{2f_{mmW}(n_\Omega+n_g)}$. In practice, one should keep the transmission line's length always to be lower than the coherence length related to the interaction between right-traveling optical mode and right-traveling mmWave mode $L_{coh}^+ = \frac{c_0}{2f_{mmW}|n_\Omega-n_g|}$. In the figure \ref{SimLengths} we present the simulated lengths for the design of the transmission line used in this work. One may see that $L_a > L_{coh}^-$ at the entire frequency range, which means that in practice, the presence of the left-traveling mmWave mode can be neglected for the length where $L_{TL} > L_a$ for two reasons: (i) the amplitude of the left-traveling wave is weak with respect to the right-traveling wave, and (ii) it is not phase-matched.

\subsection{Long transmission line case: antenna-coupled electro-optic modulator's performance}
As described above, we discuss the case of the long transmission line $L_{TL} > L_a$. In this case, the term with $A_\Omega^-$ in eq. \ref{SolutionNotneglected} might be neglected, and therefore the amplitude of the generated sideband becomes:
\begin{equation} \label{Solution}
    {A_{SB^+} = \frac{\chi^{(2)} \omega_{SB^+} }{2n_{o}c_0} A_p  \Gamma_{eo} L_{TL} \cdot A^+_{\Omega} \frac{e^{i \Delta \Tilde{k}_+ L_{TL} } - 1}{\Delta \Tilde{k}_+ L_{TL}} }     
\end{equation}
We introduce the sideband ratio ($SBR$) characterizing the efficiency of the electro-optic modulation and defined as:
\begin{equation}
    { P_{SB^+} = SBR \cdot P_p}
\end{equation}
where $P_{SB^+}, P_p$ are powers of the generated sideband and optical probe. Substituting eq.~\ref{ATHz} into eq.~\ref{Solution} gives the following expression for the sideband ratio:
\begin{equation}\label{SBR}
    {
    SBR = \frac{(\chi^{(2)} \cdot \omega_{SB^+})^2}{16 \cdot n_{o}^2 \cdot n_{\Omega} \cdot c_0^3 \cdot \varepsilon_0 \cdot Z_{TL} \cdot S_{\Omega}} \cdot  IAF^{2} \cdot |t_a|^{2} \cdot PM^{2} \cdot E_{inc}^2\cdot \Gamma_{eo}^2\cdot L_{TL}^2
    }
\end{equation}
where we introduced the phase-matching function $PM = \left| \frac{e^{i \Delta \Tilde{k}_+ L_{TL} } - 1}{\Delta \Tilde{k}_+ L_{TL}} \right| $ describing the phase-matching of optical and mmWave modes, as well as the loss of the mmWave mode. In the case of the long transmission line, $t_a\approx1-r_a$, and therefore the modulation efficiency will be directly affected by the impedance mismatch between antenna and transmission line. By denoting $P^{on-chip}_{mmW} = \frac{|V_0^+|^2}{2 Z_{TL}}=\frac{IAF^2 \cdot |t_a|^2 E_{inc}^2}{8 Z_{TL}} = \eta P^{free-space}_{mmW}$ we get:
\begin{equation}\label{SBRfinal}
    SBR = \frac{(\chi^{(2)} \cdot \omega_{o})^2\cdot \Gamma_{eo}^2\cdot L_{TL} ^2\cdot PM^{2}}{2 \cdot n_{o}^2 \cdot n_{\Omega} \cdot c_0^3 \cdot \varepsilon_0 \cdot S_{\Omega}} \cdot \eta P^{free-space}_{mmW}
\end{equation}
where we introduced the coupling efficiency coefficient $\eta$ of mmWave radiation from free-space to the transmission line and took into account that the probe frequency $\omega_o = \omega_p \approx \omega_{SB^+}$. This expression describes the modulation efficiency and might be used to experimentally calculate the coupling efficiency $\eta$ by measuring the ratio $SBR/P^{free-space}_{mmW}$.

\subsection{Short transmission line case}
In the case when $L_{TL} \leq L_a$, the right-traveling mmWave mode amplitude cannot be neglected anymore. This leads to a significant dependence of the term $t_a$ on frequency, leading to a resonant behavior of the electro-optic modulator. However, if $L_{TL}\gg L_{coh}^-$, its effect on modulating the optical probe can be neglected. Further, we discuss both of these phenomena and show that when $L_{TL} \approx L_{coh}^-$, one may develop a technique to measure the properties of the transmission line (namely, the loss coefficient $\alpha_\Omega$ and $n_\Omega$). 
\subsubsection{mmWave on-chip cavities}
First, we want to explore the resonant behaviour of the electro-optic modulators, so the transmission line terminated with a dipole antenna form a mmWave cavity. In this section, we derive the expressions for the resonant frequencies and the loss rates (both external and internal). The mmWave cavity loss rates and resonant frequency might be determined from the frequency response term $FR=\frac{|t_a|^2}{|1-r_a|^2}$:
\begin{equation}
     FR = \frac{1}{\left| 1 - r_a \cdot a \cdot e^{i2k_\Omega L_{TL} }\right|^2} = \frac{1}{\left| 1 - r_a \cdot e^{-\alpha_\Omega L_{TL}} \cdot e^{i4\pi f_{mmW} n_\Omega L_{TL} /c_0 }\right|^2}
\end{equation}

which might be rewritten as:
\begin{equation}
     FR = \frac{1}{\left| 1 - e^{i D f_{mmW} + i\theta - A} \right|^2} = \frac{1}{\left| 1 - e^{i (D f_{mmW} +\theta + iA)} \right|^2}
\end{equation}
where coefficients $D=\frac{4\pi n_\Omega L_{TL}}{c_0}$, $\theta = arg(r_a)$, $A = \alpha_\Omega L_{TL} - ln(|r_a|)$. One may find complex mmWave resonant frequency $\Tilde{f}_{mmW}^{res_m}$, satisfying condition $D\cdot \Tilde{f}_{mmW}^{res_m} + \theta + iA = 2\pi m$, where the real part $f_m = Re(\Tilde{f}_{mmW}^{res_m}) = \frac{2\pi m - \theta}{D}$ gives the resonant frequency and the resonance linewidth gives the loss rate $\kappa_{mmW}/2\pi = 2 \cdot Im(\Tilde{f}_{mmW}^{res_m}) = \frac{2A}{D}$. This can be proven as follows: introducing mmWave frequency resonance detuning $\Delta_{mmW} = 2\pi (f_{mmW} - f_m)$ and substituting it into the expression above gives:
\begin{equation}
     FR = \frac{1}{\left| 1 - e^{i(2\pi m + iA) + i\frac{D}{2\pi}\Delta_{mmW} } \right|^2} = \frac{1}{\left| 1 - e^{i\frac{D}{2\pi}\Delta_{mmW} - A} \right|^2}
\end{equation}
At small detunings (when $|i\frac{D}{2\pi} \Delta_{mmW} - A|\ll 1$) one may apply the Taylor expansion of the exponential term $e^z \approx 1 + z$, leading to:
\begin{equation}
     FR = \frac{1}{\left| i\frac{D}{2\pi}\Delta_{mmW} - A \right|^2} = \frac{4\pi^2}{D^2} \cdot \frac{1}{\left| \Delta_{mmW} + i \frac{2\pi A}{D}\right|^2} = \frac{4\pi^2}{D^2} \cdot \frac{1}{\left| \Delta_{mmW} + i \frac{\kappa_{mmW}}{2}\right|^2}
\end{equation}
which is a standard Lorentzian shape of the resonance with a linewidth $\kappa_{mmW}/2\pi = \frac{2A}{D}$. 
Substituting the expressions for coefficients $A$ and $D$ to the formula for the loss rate gives:
\begin{equation}
     \kappa_{mmW}/2\pi = c_0 \frac{\alpha_\Omega L_{TL}-ln|r_a|}{2 \pi n_\Omega L_{TL}} = \frac{1}{2\pi}\frac{\alpha_\Omega c_0}{n_\Omega} + \frac{1}{2\pi}\frac{c_0 |ln|r_a||}{n_\Omega L_{TL}} = \kappa_{mmW, in}/2\pi + \kappa_{mmW, ex}/2\pi
\end{equation}
where we split the two contributions to the loss rate - internal $\kappa_{mmW, in}$ (related to the propagation loss of the mmWave in the cavity) and external $\kappa_{mmW, ex}$ related to the loss rate induced by the presence of the antenna. 

\subsubsection{Forward and backward optical pumping schemes for photonics-enabled characterization of mmWave transmission lines}\label{Photonics_enabled_SI}

As it was described above, there are two waves propagating in the transmission line with the voltage amplitudes of $V_0^+$ and $V_0^-$, according to eq. \ref{V_def}. These voltage amplitudes correspond to the forward propagating electric field wave with an amplitude $A_\Omega$ and backward propagating wave $A_\Omega^-$. In the case of the short transmission lines, whose length $L_{TL}$ are comparable to the coherence lengths of the backward propagating wave $\frac{\pi}{\left| \Delta \Tilde{k}_{-} \right| }$, the contribution of the second term in eq.\ref{SolutionNotneglected} is not negligible. Therefore, the electric field amplitude of the generated sideband $A_{SB^{+}}^{co}$, when the optical probe co-propagates with the forward propagating mmWave mode, in accordance with eq.\ref{SolutionNotneglected} and \ref{refl_def} is proportional to:
\begin{equation}
    A_{SB^{+}}^{co} \propto \left(  \frac{e^{i \Delta \Tilde{k}_+ L_{TL} } - 1}{\Delta \Tilde{k}_+ L_{TL}} + e^{i 2 k_\Omega L_{TL} - \alpha_\Omega L_{TL}}  \frac{e^{i \Delta \Tilde{k}_{-} L_{TL} } - 1}{\Delta \Tilde{k}_{-} L_{TL}}\right)
\end{equation}
If the optical probe is sent the other way around, so that it counter-propagates the forward propagating mmWave mode, then its amplitude will be given by 
\begin{equation}
    A_{SB^{+}}^{counter} \propto \left(  \frac{e^{-i \Delta \Tilde{k}_- L_{TL} } - 1}{\Delta \Tilde{k}_- L_{TL}} + e^{i 2 k_\Omega L_{TL} - \alpha_\Omega L_{TL}}  \frac{e^{-i \Delta \Tilde{k}_{+} L_{TL} } - 1}{\Delta \Tilde{k}_{+} L_{TL}}\right)
\end{equation}
as now the phase-matching conditions will be the opposite for the forward- and backward-propagating mmWave waves. Now, by introducing the difference sideband ratio $dSBR = 10\cdot log_{10} \left( \frac{\left|A_{SB^{+}}^{co}\right|^2}{\left|A_{SB^{+}}^{counter}\right|^2} \right)$, we get:
\begin{equation}
    dSBR = 20 \cdot \log_{10} \left( 
    \frac{ 
    \left| 
    \frac{e^{i \Delta \Tilde{k}_{+} L_{TL}} - 1}{\Delta \Tilde{k}_{+} L_{TL}} 
    + 
    e^{i 2 k_\Omega L_{TL} - \alpha_\Omega L_{TL}} 
    \cdot 
    \frac{e^{i \Delta \Tilde{k}_{-} L_{TL}} - 1}{\Delta \Tilde{k}_{-} L_{TL}} 
    \right| 
    }
    { 
    \left| 
    \frac{e^{-i \Delta \Tilde{k}_{-} L_{TL}} - 1}{\Delta \Tilde{k}_{-} L_{TL}} 
    + 
    e^{i 2 k_\Omega L_{TL} - \alpha_\Omega L_{TL}} 
    \cdot 
    \frac{e^{-i \Delta \Tilde{k}_{+} L_{TL}} - 1}{\Delta \Tilde{k}_{+} L_{TL}} 
    \right| 
    }
    \right)
\end{equation}

One may extract the parameters $\alpha_\Omega$ and $n_\Omega$ by fitting the experimentally measured $dSBR$, assuming certain scaling of the losses and refractive index with $f_{mmW}$, which we discuss below in section \ref{SIlossmodel}.

\section{Modeling mmWave losses in photonics-integrated transmission lines}
\label{SIlossmodel}
In general, mmWave transmission lines are affected by  conductor and radiative losses, $\alpha_c$ and $\alpha_{rad}$, respectively. The analytical formulas for the various loss terms can be found in~\cite{keil1992high}. By introducing the frequency-dependent skin depth $\delta=\frac{1}{\sqrt{\sigma \cdot f_{mmW} \cdot \mu \cdot \pi}}$ and surface impedance $Z_S = \frac{1+i}{\sigma \delta} coth\left( (1+i) \frac{h_{Au}}{\delta} \right)$, the conductive losses are given by:
\begin{equation}
    \alpha_c = g\cdot Re \left( \frac{Z_s}{Z_{TL}} \right) \ dB/m
\end{equation}
where the expression for the proportionality coefficient $g$ is given by:
\begin{equation}
    g = 17.34 \cdot \frac{P'}{\pi w} \cdot \left(1 + \frac{w}{w_{TL}} \right) \cdot \frac{\frac{1.25}{\pi} ln(\frac{4\pi w_{TL}}{h_{Au}}) + 1 + \frac{1.25 h_{Au}}{\pi w_{TL}}}{\left( 1 + \frac{2 w_{TL}}{w} + \frac{1.25 \pi h_{Au}}{\pi w}\cdot \left( 1 + ln(\frac{4\pi w_{TL}}{h_{Au}}) \right)\right)^2}
\end{equation}
where $w, w_{TL}, h_{Au}$ are the parameters as presented in section \ref{SIstructure}, and $P'$ is:
\begin{equation}
    P' = 
\begin{cases}
\frac{k}{(1 - \sqrt{1-k^2})(1-k^2)^{3/4}}  \left( \frac{K(k)}{K'(k)}\right)^2, & \text{for } 0\leq k\leq0.707 \\
\frac{1}{(1-k)\sqrt{k}}, & \text{for } 0.707\leq k\leq 1
\end{cases}
\end{equation}
with $k=\frac{w}{w + 2w_{TL}}$, $K(k)$ being a complete elliptic integral of the first kind, and $K'(k) = K(\sqrt{1-k^2})$. As shown in the same work, the radiative losses are proportional to the third power of frequency:
\begin{equation}
    \alpha_{rad} = \pi^5 \frac{3 - \sqrt{8}}{2} \frac{n_\Omega}{n_{sub}}\cdot \left(1 - \frac{n_\Omega^2}{n_{sub}^2} \right)^2 \cdot \frac{(w+2w_{TL})^2}{c_0^3 K'(k) K(k)} f_{mmW}^3
\end{equation}
where $n_{sub}\approx3.41$ is the substrate's refractive index (silicon). Another contribution to the loss, a tangent loss $\alpha_t$, can be simulated using, for example, CST Studio Software by simulating the 2D mode profile and relating the imaginary part of the effective refractive index $\kappa_\Omega$ as:
\begin{equation}
    \alpha_{t} = \frac{4\pi \kappa_\Omega f_{mmW}}{c_0}
\end{equation}

\begin{figure}[h!]
    \centering
    \includegraphics[width=0.8\linewidth]{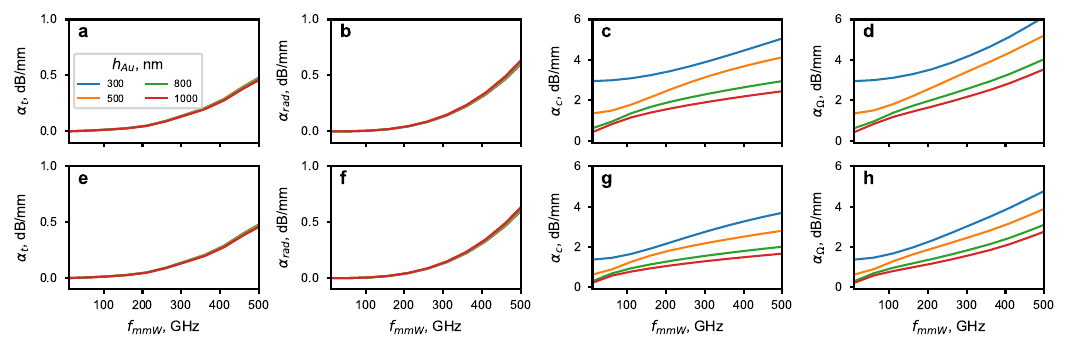}
    \caption{\textbf{Simulated loss for various electrode thicknesses and metal properties.} (a, e) Tangent, (b, f) radiative, (c, g) conductor and (d, h) total loss versus mmWave frequency. Upper row: $\sigma = 1.91 \cdot 10^7$ S/m (as retrieved from the experimental data). Lower row: $\sigma = 4.1 \cdot 10^7$ S/m (literature value of gold conductance).}
    \label{SimAbs}
\end{figure}

Combining all the above mentioned contributions of the loss we define the total loss $\alpha_\Omega = \alpha_c + \alpha_{rad} + \alpha_t$ and plot these various contributions in Fig.~\ref{SimAbs}. As it is clearly seen, the most dominant contribution comes from the conductor losses. One of the strategies on how to minimize this loss is to increase the metal thickness $h_{Au}$. We plot the losses for the various metal thicknesses $h_{Au}$ in Fig.\ref{SimAbs}. One may notice that the losses can be reduced significantly by increasing the metal thickness. The fabrication of the thicker electrodes might be challenging, however, it has been already demonstrated in~\cite{wang2024ultrabroadband}.

\section{Measurements of the antenna-transmission line reflection coefficient}\label{ReflmodelSI}

\begin{figure}[h!]
    \centering
    \includegraphics[width=0.8\linewidth]{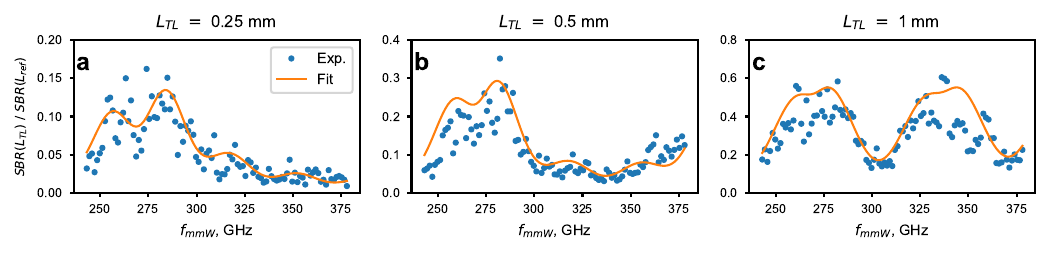}
    \caption{The sideband ratio as a function of mmWave frequency for the devices with the length of transmission line (a) 0.25 mm; (b) 0.5 mm; (c) 1 mm.}
    \label{figReflCoefMeas300GHz}
\end{figure}

Guided by the theory developed in the previous section, we now experimentally confirm the impedance mismatch between the antenna and its feeding transmission line by measuring the complex reflection coefficient $r_a$ at their junction. Together with the near-unity reflection of the open circuit at the end of the transmission line, the reflection at the antenna-transmission line interface supports the formation of a mmWave cavity that will shape the $SBR$ as a function of mmWave frequency. According to our theoretical model, neglecting the left-traveling  mmWave mode, the dependence of the $SBR$ is proportional to (eq. \ref{SBR}):
\begin{equation}
    SBR \propto  PM^2 \cdot L_{TL}^2 \cdot |t_a|^2 
\end{equation}
where we grouped all terms with dependence on $L_{TL}$. Using extracted mmWave losses $\alpha_\Omega$ and refractive index $n_\Omega$ with a technique described in \ref{Photonics_enabled_SI}, the ratio of $SBR$ at two lengths $L_{TL}$ and $L_{ref}$ is given by:
\begin{equation}\label{SBR_for_fit}
    \frac{SBR(L_{TL})}{SBR(L_{ref})} = \frac{PM(L_{TL})^2}{PM(L_{ref})^2} \cdot \frac{L_{TL}^2}{L_{ref}^2} \cdot \frac{\left|1 - r_{a} \cdot a \cdot e^{i 2 k_{\Omega} L_{ref}}\right|^2}{\left|1 - r_{a} \cdot a \cdot e^{i 2 k_{\Omega} L_{TL}}\right|^2} 
\end{equation}
where only one unknown parameter left: reflection coefficient $r_a$ which can be extracted by fitting the frequency dependence of $\frac{SBR(L_{TL})}{SBR(L_{ref})}$ for various lengths of the transmission line. 
We measure $SBR$ for four devices with lengths of 0.25, 0.5, 1 and 2 mm, and we use 2 mm long device as a reference ($L_{ref}=2$ mm) and plot the spectra of $\frac{SBR(L_{TL})}{SBR(L_{ref})}$ in Fig.\ref{figReflCoefMeas300GHz} and their fits using eq. \ref{SBR_for_fit}, where we neglect the frequency dependence of $r_a$, since simulations show relatively small variation of both phase and amplitude of $r_a$ within the given frequency range.

The fitted value of the $r_a^{fit}=0.61e^{-i\cdot0.041\pi}$ is very close to the simulated one plotted in Fig.\ref{fig3} of the main text both in amplitude and in phase. 

\begin{figure}[h!]
    \centering
    \includegraphics[width=0.53\linewidth]{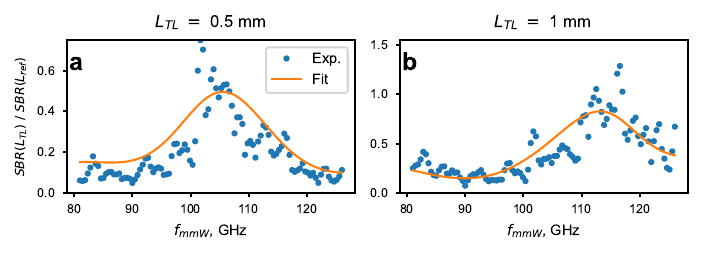}
    \caption{The sideband ratio as a function of mmWave frequency for the devices with the length of transmission line (a) 0.5 mm; (b) 1 mm.}
    \label{figReflCoefMeas100GHz}
\end{figure}

Then, we perform the same measurements across the WR9.0 band, and plot the results in Fig.\ref{figReflCoefMeas100GHz}. The fitted value of the $r_a^{fit}=0.69e^{i\cdot0.34\pi}$ is again very close to the simulated one plotted Fig.\ref{fig3} of the main text.

\section{Supporting measurements for the electro-optic comb generation experiments}
\subsection{Linewidth of the optical microring resonance}\label{LwSI}
\begin{figure}[h!]
    \centering
    \includegraphics[width=0.53\linewidth]{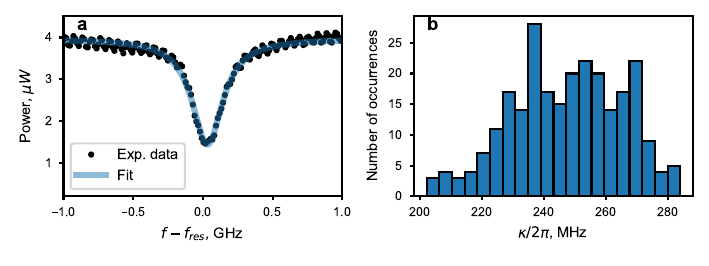}
    \caption{(a) The transmission spectrum of the pumped resonance and (b) linewidth statistics over 256 resonances}
    \label{figOptRes}
\end{figure}

We performed the transmission measurement of the ring resonator in order to get the linewidth of the pumped resonance. We plot the transmission spectrum of the resonance we pumped for the experiments in the electro-optic comb generation in Fig.\ref{figOptRes} as well with the statistics over multiple resonances. We pumped the resonance with the linewidth of 220 MHz and this value was further used in the calculations.

\subsection{Measurements of the mmWave coupling efficiency for the frequency comb experiments}\label{CouplEffRingsSI}
As described above, we measured the coupling efficiency each day prior to conducting the main experiments. For these measurements, we used a test device consisting of a 2mm-long transmission line with a dipole antenna. The measured coupling efficiencies are presented below in Fig.\ref{figCoupleffRings}. Using the linear fit, we extracted the coupling efficiency at 123.2 and 307.9 GHz to be 0.027 and 0.045, respectively. 

\begin{figure}[h!]
    \centering
    \includegraphics[width=0.4\linewidth]{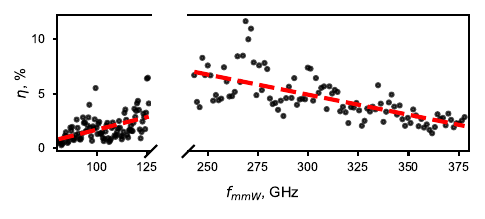}
    \caption{Coupling efficiency of the mmWave free-space radiation to the chip performed right before electro-optic comb measurements.}
    \label{figCoupleffRings}
\end{figure}

We note the coupling efficiency at frequencies below 125 GHz are lower than the previous experiments because the TPX lens could not be used due to the setup's constraints and the limited space behind the chip. 

\subsection{Calculation of the loss rates of the mmWave cavity}\label{LossRatesSI}

Using the result of the measurement of the loss rate $\alpha_\Omega$ and the reflection coefficient $r_a$ in the main text Fig. \ref{fig3}, we calculated the total, internal and external loss rates as well as the closest resonant frequencies for $L_{TL}$=1.5mm for the experimental mmWave frequencies of 123.2 and 307.9 GHz and present them in Table \ref{tab:lossrates}. We also estimated the Q factors (total and internal) of the mmWave cavity.

\begin{table} [h!]
    \centering
    \begin{tabular}{cccccccc}
         $f_{mmW}$ & $\kappa_{mmW}/2\pi$ & $\kappa_{mmW, in}/2\pi$ & $\kappa_{mmW, ex}/2\pi$ & $f^{res_m}_{mmW}$ & $\Delta_{mmW}/2\pi$ & $Q_{mmW, in}$ & $Q_{mmW}$
         \\
         123.2 GHz & 20.67 GHz & 15.54 GHz & 5.13 GHz & 122.26 GHz & 0.94 GHz & 7.87 & 5.96
         \\
         307.9 GHz & 26.51 GHz & 19.67 GHz & 6.84 GHz & 315.05 GHz & 7.15 GHz & 15.65 & 11.61
         \\
    \end{tabular}
    \caption{Parameters of the mmWave cavity used for the frequency comb experiments}
    \label{tab:lossrates}
\end{table}

\subsection{Measurements of $g_{eo}$ from optical resonance splitting}\label{ResonanceSplitting}

\begin{figure}[h!]
    \centering
    \includegraphics[width=0.6\linewidth]{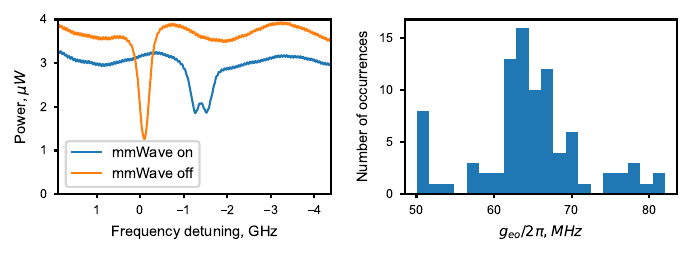}
    \caption{\textbf{Resonance splitting experiments.} (a) Measurements of the microring transmission spectrum with and without mmWave radiation. The mmWave power was set to 18.7 dBm, and the mmWave frequency is 123.2 GHz. (b) A histogram showing the statistics over multiple resonances.}
    \label{ResonanceSplittingFig}
\end{figure}

We measured the splitting of the resonance under illumination with mmWave radiation as was suggested in \cite{zhang2025ultrabroadband}. When mmWave is illuminating the microring resonator, the microring resonance splits into two resonances, with a distance between peaks equal to $g_{eo}/4$, as we show in Fig. \ref{ResonanceSplittingFig}(a). We analyzed the statistics over multiple optical resonances and found the mean $g_{eo}/2\pi = 64.7 \pm 7.0 $ MHz, consistent with the values extracted from the comb's slope, resulting in the same $g_0/2\pi$ of $4.79 \pm 0.52$ kHz. 

\section{Strategies on improving the coupling efficiency of the mmWave radiation to the chip}\label{CouplEffStrategiesSI}

\begin{figure}[h!]
    \centering
    \includegraphics[width=0.8\linewidth]{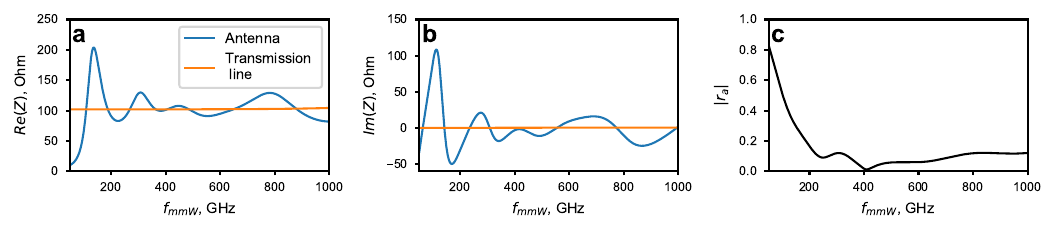}
    \caption{\textbf{Simulation results for the bow-tie antenna.} (a) Real part of the impedance; (b) Imaginary part of the impedance; (c) reflection coefficient between the bow-tie antenna and the transmission line demonstrating flat and low reflection coefficient $r_a$ across frequency range of 200-1000 GHz.}
    \label{ReflBowtie}
\end{figure}

As discussed above, the coupling efficiency depends mainly on two parameters: $IAF$ and $r_a$. Then, $IAF$ might be optimized in two different ways: by optimizing the parameters of the free-space gaussian mmWave beam (such as the beam waist) illuminating the chip by developing the lens system, or by engineering the design of the antenna and using other types of antennas, such as bow-tie~\cite{benea2018three}, bull eye~\cite{6661458}, LC~\cite{salamin2019compact} and others. 
In addition, the reflection coefficient at the antenna-transmission line $r_a$ significantly affects the coupling efficiency. Broad- or narrowband coupling efficiency optimization is considered depending on the application. In the first case, the $r_a$ must be minimized across the entire frequency range of interest, bringing the values of $t_a = \frac{1-r_a}{1 - r_a \cdot a\cdot e^{i2k_\Omega L}}$ to 1. However, in the case of narrowband optimization, one may consider exploiting the resonant behavior of the term $t_a$ as $r_a  \rightarrow -1$. By minimizing the round-trip propagation loss $a$, one may benefit from the enhancement of the intra-cavity field at frequency where $arg(r_a \cdot e^{i2k_\Omega L}) = 2\pi m$ with $m$ being an integer number. 
We demonstrate the possibility of engineering the $r_a$ coefficient to be significantly low across a large frequency range by simulating the reflection coefficient of the bow-tie antenna with a width of 300 $\mu$m and a height of 400 $\mu$m, as presented in the Fig. \ref{ReflBowtie}. 

\listoffigures

\end{document}